	   \documentclass[oldversion]{aa}     
           \usepackage{txfonts}
	   \usepackage{epsf}
           \usepackage{amssymb}
           \newcommand{\ud}{\mathrm{d}}

\begin{document}

   \title{A consistent solution for the velocity field and mass-loss rate of massive stars}

   \author{Patrick~E.~M\"uller\inst{1} \and Jorick~S.~Vink\inst{2}}
 
   \offprints{P.~E.~M\"uller, \email{pmueller@astro.keele.ac.uk}}

   \institute{School of Physical and Geographical Sciences,
              Lennard-Jones Laboratories,
              Keele University,
              Staffordshire,
              ST5 5BG,
              UK
	      \and
	      Armagh Observatory,
              College Hill,
              Armagh, BT61 9DG,
              Northern Ireland,
	      UK}
   
     \date{Accepted 7 October 2008}

 \abstract
{Stellar winds are an important aspect of our understanding of the evolution of massive stars and their input into the interstellar medium.
Here we present solutions for the velocity field and mass-loss rates for stellar outflows as well as for 
the case of mass accretion through the use of the so-called Lambert W-function. For the case of a radiation-driven wind, the 
velocity field is obtained analytically using 
a parameterised description for the line acceleration that only depends on radius, which we obtain 
from Monte-Carlo multi-line radiative transfer calculations.
In our form of the equation of motion the critical point is the sonic point.
We also derive an approximate analytical solution for the supersonic flow which closely resembles 
our exact solution. 
For the simultaneous solution of the mass-loss rate and velocity field, we describe a new iterative method. We apply our 
theoretical expressions and our iterative method to the stellar wind from a typical O5--V main sequence star, and find 
good agreement with empirical values. 
Our computations represent a self-consistent mass-loss
calculation including the effect of multi-line scattering for an O--type star, opening up the possibility of applying 
Monte Carlo mass-loss calculations in regions of the Universe for which empirical constraints cannot be readily obtained.
    }

  \titlerunning{Radiation-driven winds from massive stars}

  \maketitle

\section{Introduction}

Stellar winds are crucial for our understanding of the evolution of massive stars towards their final explosion.
Furthermore, mass loss constitutes an important source of chemical enrichment into the interstellar medium. 
As we cannot obtain observations of individual stars at high redshift, we need to develop a theoretical framework 
that works locally and which confidently can be applied to the 
early Universe. Developing such a framework is the aim of this paper. 

That the winds from massive O stars are driven by radiation pressure through spectral lines has been 
known since the seminal papers of Lucy \& Solomon (1970) and Castor, Abbott \& Klein (\cite{castor:abbott:klein}, 
hereafter CAK).
Currently, there are two basic methods in use to compute the mass-loss rates of massive stars. 
These involve a modified version of the CAK method (Pauldrach et al. \cite{pauldrach:al}, Kudritzki \cite{kudritzki}) 
and the Monte Carlo method (Abbott \& Lucy \cite{abbott:lucy}, Vink et al. 2000). 
The {\it pros} and {\it cons} of these two distinct approaches have recently 
been reviewed by Vink (2006).

A common feature of supersonic flows are critical points. 
Within the CAK framework, the wind velocity, $v(r)$, has critical points defined by 
three singularity conditions, 
whilst assuming 
a power-law for the line acceleration, 
to obtain an approximate solution of the equation of motion.   
To solve this equation, CAK attributed the same status to the Sobolev velocity derivative as to the Newtonian 
derivative. This CAK {\it ansatz} was immediately 
criticised by Lucy (\cite{lucy}, \cite{lucy98}, \cite{lucy07b}, see also Vink et al. \cite{vink:dekoter:lamers}, Vink \cite{vink}), 
but the CAK-scaling relations have been widely used, presumably because 
of their reasonable agreement with observations. The analyses of continuum emission and 
line profiles yielded mass-loss rates $\dot{\cal M}$ from 10$^{-6}$ to \mbox{10$^{-5} {\cal M}_{\sun}$/year}, 
and terminal velocities of about three times the escape velocity, $\sim$3000 km/s for a main-sequence O-star
(e.g. Howarth \& Prinja \cite{howarth:prinja}). 
That the CAK predictions are in rough agreement with observations at all is largely fortuitous (see Vink \cite{vink06}), 
as the assumptions of local thermodynamic equilibrium and the early use of limited line lists are {\it a posteriori} known
to be insufficient. The CAK framework itself however has proved to be very successful and improvements 
in non-LTE radiative transfer as well as atomic data have led to more reliable predictions by e.g. Pauldrach et al. (1986). 
Nevertheless, even these modified CAK scaling relations have not been able to reproduce 
the mass-loss rates of the denser O star winds (Lamers \& Leitherer \cite{lamers:leitherer}, Vink et al. \cite{vink:al:00}), nor 
the winds of Wolf-Rayet stars (Lucy \& Abbott \cite{lucy:abbott}, Gayley \cite{gayley}, Gr\"afener et al. \cite{graefener:al},
Gr\"afener \& Hamann \cite{graefener:hamann}).

For this reason, an alternative method for predicting mass-loss rates, the Monte Carlo (MC) technique, was developed 
(Abbott \& Lucy \cite{abbott:lucy}, Puls \cite{puls}, Schmutz \cite{schmutz}, de Koter et al. \cite{dekoter:heap}, Vink et al. \cite{vink:dekoter:lamers}).
The MC approach does not rely on the CAK ansatz, whilst still employing the computational benefits of 
the Sobolev approximation in radiative transfer. In the MC-method, 
mass loss does not constitute an eigenvalue of the differential equations that govern the stellar wind, 
but the rates follow from assuming a wind velocity structure -- in the form of a so-called 
$\beta$-law (Castor \& Lamers \cite{castor79}, CAK). The advantages of using the MC-method are manifold. 
There is no dependence on the CAK ansatz and the resulting CAK critical point,
whilst the line acceleration is computed at all points in the wind without relying 
on a simplistic force multiplier approach involving a 2-parameter power-law approximation 
for the line acceleration. Furthermore, multiple scatterings 
can be included naturally through Monte Carlo simulations. 
Because the kinetic energy density at the sonic point is negligible compared to that in the terminal flow, the 
MC-method does not rely on details of the trans-sonic flow (Lucy \cite{lucy07a}). 
Finally, the terminal wind velocity is a direct and accurate (within $\sim$10\%) observable quantity 
in the local Universe. 

It is probably for these reasons that the Vink et al. (\cite{vink:al:00}) MC 
mass-loss rates\footnote{http://star.arm.ac.uk/$\sim$jsv/} are widely used in evolutionary 
calculations, but however successful the MC approach 
may be in correctly predicting the mass-loss metallicity dependence (Vink et al. \cite{vink:al:01}, Mokiem et al. \cite{mokiem:al:07b}),
the current MC approach is semi-empirical. Just as we are not able to obtain empirical mass-loss rates from the early Universe, 
neither will we be able to obtain direct wind velocities.
It is therefore important that we develop a reliable theoretical 
framework through which we are able to predict the mass-loss rate and wind velocity simultaneously.

We first derive analytical solutions for the general velocity and density structure 
in any mass outflow or inflow situation. 
The solutions are obtained using the so-called Lambert W function (M\"uller \cite{mueller}, Cranmer \cite{cranmer}).
As our solutions are presented in an explicit analytical form, physical insight can more easily be obtained through
parameter changes. This in contrast to pure numerical models, where it is more challenging to predict the 
response of changes in the underlying parameters. For the specific case of a radiation-driven wind, we do not rely on the 
CAK expression for the radiation force, rather we describe the line acceleration as a function of 
stellar radius, $g(r)$. In addition, the critical point of our stellar wind is the sonic point 
(as in Parker \cite{parker} solar wind theory), and not the CAK critical point. 
The calculation of $g(r)$ is performed through Monte Carlo simulations accounting for multi-line transfer, 
and the wind parameters are solved simultaneously in an iterative way. 

The set-up of the paper is as follows.
In Sects.\ref{sect-basic-hydro-eq}--\ref{sect-line-acc-term}, the hydrodynamic equations for a spherically symmetric steady flow
are introduced including a derivation of the mathematical description of the radiative line acceleration as a function of
radius for the case of a radiation-driven wind.
The process for obtaining the fully analytical 1D solutions is described and discussed in Sect.~\ref{sect-analyt-sol-Eq-mot}. 
Here, the velocity field for the entire family of solutions is provided in an explicit general expression
from which the solutions for a radiation-driven wind or mass accretion flux (e.g. collapsing protostellar cloud)
follow as unique trans-sonic solutions through the critical point.
Moreover, an approximate analytical solution for the supersonic flow is presented.
In Sect.~\ref{sect-num-meth}, we describe our numerical computation obtaining the radiative acceleration in our stellar wind models. 
Furthermore, a new iterative method for the determination of the consistent solution
for the mass-loss rate is provided.
In Sect.~\ref{sect-applic-ostar}, we present the application of our models to the stellar wind from a typical O5--V-star, 
and discuss the results, before we summarise and discuss our findings in Sect.~\ref{sect-disc}.

 \section{Radiation hydrodynamics of expanding or collapsing systems}  \label{sect-analytical-theory}

 \subsection{Basic equations of hydrodynamics} \label{sect-basic-hydro-eq}

      Considering a non-viscous, i.e. ideal fluid, the 
      momentum equation 
          \begin{equation} \label{momentum-eq}
            \rho\, \frac{{\rm D}\,\vec{v}}{{\rm D}\,t} = \vec{f} - \vec{\nabla}\,p
          \end{equation} 
      is valid (see, e.g., Mihalas \& Weibel Mihalas \cite{mihalas:mihalas}), 
      where $D/D\,t$ is the covariant Lagrangean or co-moving time derivative in
      the fluid-frame of a material element and \vec{v} is its velocity, 
      $\vec{f}$ is the total external body force per volume acting on a mass element of fluid,   
      and $\vec{\nabla}\,p$ is the divergence term of a diagonal isotropic stress tensor
      $\vec{\nabla}\cdot{}\tens{T}$, in which $\tens{T}=-p\,\tens{I}$ and $p$ is the hydrostatic pressure. 
 
      One also needs to consider the
      equation of continuity
              \begin{equation} \label{eq-continuity}
                    \frac{\partial\rho}{\partial{}t} + \vec{\nabla}\cdot{}\left( \rho\,\vec{v}\right) = 0 \, ,
                \end{equation} 
     with the covariant divergence $\vec{\nabla}\cdot{}\vec{v}$ of the velocity vector.

    \subsection{Simplifying assumptions} \label{sect-simpl-assump}
 
     In order to solve the hydrodynamic equations analytically, and to 
     account for the spherically symmetric stationary problem, we make the following
     simplifying assumptions: 
     \begin{enumerate}
       \item The stellar wind is isothermal with a temperature equal to the effective temperature $T_{\rm eff}$
                       of the central star.
                       In this case, the equation of state
                       \begin{equation}   
                       p = a^{2}\,\rho 
                      \end{equation} 
                      is valid, where $a$ is the isothermal speed of sound and
                      $\rho$ is the density of the wind.   
       \item We assume a stationary 1D spherical flow,
                       \begin{equation}    
                       \frac{\partial}{\partial{}t} \equiv 0, 
		       \quad \frac{\partial}{\partial{}\phi} = \frac{\partial}{\partial{}\theta}  \equiv 0, \quad 
                       f_{\rm\phi} =  f_{\rm\theta}  \equiv 0
                     \end{equation}  
                      i.e.~no shocks, no clumps, and no significant distortion of the star.
	\item In the case of a wind from a luminous early-type star, the wind is primarily driven by
                      continuum plus line radiation forces,
		      where the radial acceleration on a mass element is
                       \begin{equation}  
                       \frac{f_{\rm r}}{\rho} = - \frac{G\,M}{r^{2}}\, \left(1-\Gamma \right) + g_{\rm rad}^{\rm line}
                       \end{equation}
		       with
		       \begin{equation} 
		        \Gamma := \frac{g_{\rm rad}^{\rm cont}}{g}\, ,
		       \end{equation}
		       the force ratio between the radiative acceleration $g_{\rm rad}^{\rm cont}$ due to the continuum opacity
		       (dominated by electron scattering) and the inward acceleration
		       of gravity $g$, and
		       $g_{\rm rad}^{\rm line}\,(r)$ is the outward radiative acceleration due to spectral lines.
		       $M$ is the mass of the central star.
       \end{enumerate}

  \subsection{Simplified hydrodynamic equations}
       
  \subsubsection{Wind density and mass-loss rate}

     If we use the covariant derivative 
     (see Mihalas \& Weibel Mihalas \cite{mihalas:mihalas}) 
     for spherical coordinates and apply assumption 2 to the equation
     of continuity~(\ref{eq-continuity}), we find
              \begin{equation} \label{eq-continuity2} 
                   \frac{\partial}{\partial{}r}\,\left( r^{2}\,\rho\,v \right) = 0 
                \end{equation} 
   for a one-dimensional, spherically symmetric and steady flow.
   Integrating this equation yields the conserved total mass flux (or mass-loss rate)   
   through a spherical shell with radius $r$ surrounding the star
          \begin{equation}  \label{Mdot}
       \dot{M} =  4\,\pi\, r^{2}\, \rho\,(r)\,  v\,(r) = \mbox{constant}\, ,   
           \end{equation} 
  or equivalently, for the wind density
        \begin{equation} \label{rho}  
     \rho\,(r) = \frac{\dot{M}}{4\,\pi\,r^{2}\, v\,(r)}
                                    = \frac{F}{{\hat r}^{2}\, v\,({\hat r})} 
         \end{equation}   
   with the defined flux \mbox{$F=\dot{M}\,/4\,\pi\,R^{2}$} through the star's surface at radius $R$
   and the dimensionless radius ${\hat r}=(r/R)$.

   Important note:\, All formulae derived in this Sect.~\ref{sect-analytical-theory} are expressed in terms of ${\hat r}$
   referring to the radius R, which is (throughout this section) the stellar (i.e. photospheric) radius
   of the central star. 
   However, all formulae can generally also be applied regarding the reference radius R to be the
   inner boundary radius $R_{\rm in}$ (close to the stellar photosphere), from where the computations of a numerical wind model start
   (see, e.g., the results for an O-star in Sect.~\ref{sect-applic-ostar} and associated diagrams in this Sect.~\ref{sect-analytical-theory}).

   \subsubsection{The equation of motion}  \label{sect-eq-of-mo}
   
     By using the correct contravariant components of acceleration $({\rm D}\,v_{i}/{\rm D}\,t)$ in Eq.~(\ref{momentum-eq}), 
     for spherical coordinates, and replacing them by their equivalent physical components
     (see again, e.g., Mihalas \& Weibel Mihalas \cite{mihalas:mihalas}),
     and applying assumptions 1-3, we obtain the simplified 
     r-component of the momentum equation of a spherically symmetric steady flow
      \begin{equation} \label{Eq-of-mot1-nd}         
          {\hat v}\,\frac{\ud}{\ud{\hat r}}\,  {\hat v}      
             =  -\frac{{\hat v_{\rm crit}}^{2}}{{\hat r}^{2}} +  {\hat g_{\rm rad}^{\rm line}}
                 -  \frac{1}{\rho}\, \frac{\ud \rho}{\ud{\hat r}} \, ,
     \end{equation}
    in dimensionless form.
    The following dimensionless velocities (in units of the isothermal sound speed $a$)
     \begin{equation} 
      {\hat v} := \frac{v}{a}\, , \quad
      {\hat v_{\rm crit}} := \frac{1}{a}\, \sqrt{\frac{G M}{R}\,(1 - \Gamma)}\, ,
     \end{equation}
     and dimensionless line acceleration
        \begin{equation}   \label{def-eq-grad-hat} 
           {\hat g_{\rm rad}^{\rm line}} := \frac{R}{a^2}\, g_{\rm rad}^{\rm line}  \, .
        \end{equation} 
    are used.
    $v_{\rm crit}$ equals the rotational break-up velocity in the case of a corresponding 
    rotating star, but here it is simply the effective escape velocity $v_{\rm esc}$ divided by
    a factor of $\sqrt{2}$
     \begin{displaymath}
     v_{\rm crit} \equiv \frac{v_{\rm esc}}{\sqrt{2}} \, .
     \end{displaymath}
    By means of Eq.~(\ref{rho}) and applying the chain rule to the function
    $1/v_{\rm r}({\hat r})$, we obtain
     \begin{eqnarray*}    
       \frac{\ud \rho}{\ud{\hat r}}  & = & 
            \left( -\frac{2}{ {\hat r}^{3} }\, \frac{1}{ v\, ({\hat r}) } - \frac{1}{{\hat r}^{2}}\, \frac{1}{{ v\,
                ({\hat r})}^{2}} \, \frac{\ud v\,({\hat r})}{\ud{\hat r}}    \right)\, F \\
         & \equiv & - \rho\,({\hat r})\, \left( \frac{2}{{\hat r}} 
                                                   + \frac{1}{v\,({\hat r})}\, 
                                                   \frac{\ud v\,({\hat r})}{\ud{\hat r}}
                                                    \right)   \, .                
     \end{eqnarray*} 
     Using this expression for \mbox{$\ud \rho / \ud{\hat r}$} in Eq.~(\ref{Eq-of-mot1-nd}), we finally find
     the dimensionless differential {\em equation of motion\/}
     for the radial velocity
         \begin{equation}  \label{Eq-of-mot2}  
         \left( {\hat v} - \frac{1}{ {\hat v} }\right)\, \frac{\ud}{\ud\,{\hat r}}\, {\hat v} =
                -\frac{{\hat v_{\rm crit}}^{2}}{{\hat r}^{2}} + \frac{2}{\hat r} 
                + {\hat g_{\rm rad}^{\rm line}} \, ,
	\end{equation}
     that is now independent of $\rho$.

    \subsection{The line acceleration term and the final equation of motion}  \label{sect-line-acc-term}

\begin{figure}
\centerline{\hspace{0cm}\epsfxsize=9.5cm \epsfbox{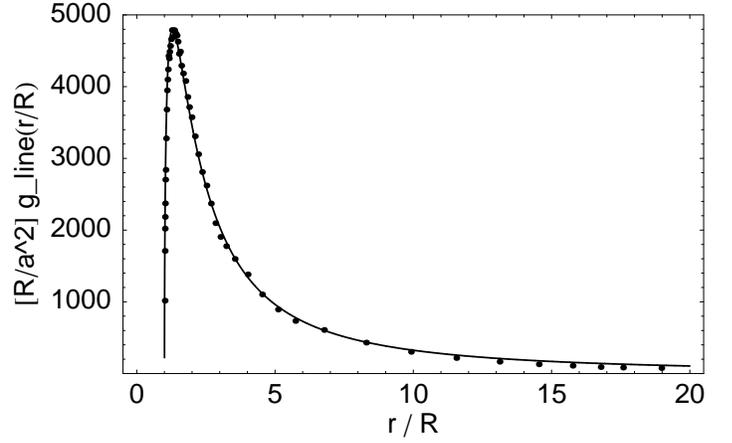}}
\caption{The dimensionless radiative line acceleration ${\hat g_{\rm rad}^{\rm line}}\,({\hat r})$ 
         vs. radial distance ${\hat r}$ (in units of $R=11.757\,R_{\sun}$) in the wind from a typical O5--V-star
	 (see stellar parameters in Sect.~\ref{sect-applic-ostar}).
	 The dots represent the results from a numerical calculation of ${\hat g_{\rm rad}^{\rm line}}\,({\hat r}_{i})$
	 for discrete radial grid points ${\hat r}_{i}$.
	 In order to determine the line acceleration parameters ${\hat g_{0}}$, $\gamma$, $\delta$ and ${\hat r_{0}}$
	 in Eq.~(\ref{line-acc-term1}), 
	 these values were fit to this non-linear model equation resulting in 
         the displayed fitting curve (solid line) with the parameters
	 ${\hat g_{0}}=17661$, $\gamma=0.4758$, $\delta{}=0.6878$ and ${\hat r_{0}}=1.0016$ (according to $v_{\infty}=3232$~km/s), 
	 at the end of the iteration process (of type A), described in Sect.~\ref{sect-it-method} and Sect.~\ref{sect-applic-ostar}, respectively.
	 }
\label{pic-gline-fit}
\end{figure}
    
    To derive a sophisticated mathematical expression for the radiative line acceleration
    as a function of radius $r$ only,
    we collect all the physically motivated mathematical properties of this function 
    \mbox{${\hat g_{\rm rad}^{\rm line}}\,({\hat r})$}:
    \begin{enumerate}
      \item For hydrostatic reasons, ${\hat g_{\rm rad}^{\rm line}}$ should be zero at a particular radius ${\hat r'}(\approx{}1)$ near the 
            stellar photosphere:
            \begin{displaymath}
              {\hat g_{\rm rad}^{\rm line}}\,({\hat r}={\hat r'}) = 0\,,
            \end{displaymath}
      \item ${\hat g_{\rm rad}^{\rm line}}$ should always be positive (except close to the stellar surface, see item 1)
            as the radiation from the photosphere is streaming outwards and so is the radiative force:
	    \begin{displaymath}
              {\hat g_{\rm rad}^{\rm line}}\,({\hat r})>0 \quad \mbox{for} \quad {\hat r}>{\hat r'} \, ,
            \end{displaymath}
      \item ${\hat g_{\rm rad}^{\rm line}}$ is supposed to decrease (i.e.~goes to zero) as \mbox{$1 / {\hat r}^2$}
            (cf. Castor \cite{castor} and Abbott \cite{abbott}) 
            with increasing radial distance ${\hat r}$ from the central star:
	    \begin{displaymath}
            {\hat g_{\rm rad}^{\rm line}}\,({\hat r}) \approx \frac{1}{{\hat r}^2} \longrightarrow 0
	    \quad \mbox{for} \quad {\hat r} \longrightarrow \infty \, ,
            \end{displaymath}
      \item ${\hat g_{\rm rad}^{\rm line}}$ has an absolute maximum somewhere in the range between the
            stellar surface and the outer edge of the wind because of the properties 1-3 above:
	    \begin{displaymath}
	  0 \leq {\hat g_{\rm rad}^{\rm line}}\,({\hat r}) \leq {\hat g_{\rm rad}^{\rm line}}\,({\hat r}_{\rm max})\, .	     
	    \end{displaymath}
    \end{enumerate}
    All these mathematical properties for the radiative line acceleration term can be realised
    by the function
     \begin{equation} \label{line-acc-term1}
        {\hat g_{\rm rad}^{\rm line}} ({\hat r}) 
         = \frac{\hat g_{0}}{\hat r^{1+\delta}}\, \left( 1 - \frac{\hat r_{0}}{{\hat r}^{\delta}} \right)^{\gamma}   \,
    \end{equation}
   or equivalently,
   \begin{equation} \label{line-acc-term2}
   {\hat g_{\rm rad}^{\rm line}} ({\hat r}) 
         = \frac{\hat g_{0}}{\hat r^{1+\delta\,\left( 1+\gamma \right)}}\, \left({{\hat r}^{\delta}} - {\hat r_{0}}\right)^{\gamma}  \, ,
   \end{equation}
   which is independent of ${\hat v}$ and \mbox{($\ud\,{\hat v} / \ud\,{\hat r}$)},
   and is a function of ${\hat r}$ only.
   
   We note that the position of the maximum \mbox{(${\hat r}_{\rm max},{\hat g_{\rm rad}^{\rm line}}\,({\hat r}_{\rm max})$)}
   has to be variable and adjustable for each set of stellar and wind parameters, as the number and species of spectral lines that make up
   the radial force are also variable with each stellar wind. We therefore had to 
   introduce the two parameters, ${\hat g_{0}}$ and $\gamma$.
   Moreover, item~1 is fulfilled at radius ${\hat r'}={\hat r_{0}}^{1 / \delta}$
   and item~3 can be guaranteed by appropriate values of the exponents $\delta$ and $\gamma$
   (where $0<\gamma\lesssim{}1$ and $0<\delta\lesssim{}1$).
   Altogether, we had to introduce at least the set of four parameters in Eqs.~(\ref{line-acc-term1}) and~(\ref{line-acc-term2}).
   
   Thus, the equation of motion~(\ref{Eq-of-mot2}) becomes 
    \begin{equation}  \label{Eq-of-mot3}  
         \left( {\hat v} - \frac{1}{ {\hat v} }\right)\, \frac{\ud}{\ud\,{\hat r}}\, {\hat v} =
                -\frac{{\hat v_{\rm crit}}^{2}}{{\hat r}^{2}} + \frac{2}{\hat r} 
                + \frac{\hat g_{0}}{\hat r^{1+\delta\,\left( 1+\gamma\right)}}\, \left({\hat r}^{\delta} - {\hat r_{0}}\right)^{\gamma}  \, .
    \end{equation}
   This equation is fully solvable analytically as we show later.

    \subsection{Analytical solutions of the equation of motion} \label{sect-analyt-sol-Eq-mot}

    \subsubsection{The critical point and critical solutions} \label{sect-crit-point-sol}

\begin{figure}
\centerline{\hspace{0cm}\epsfxsize=9.5cm \epsffile{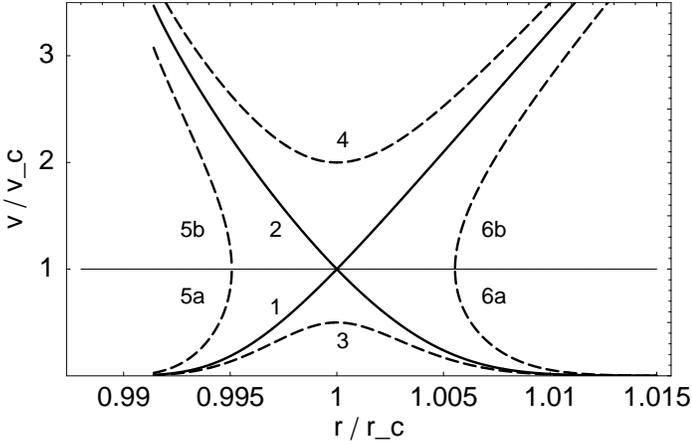}}
\caption{The topology of solutions $|v({\hat r}/{\hat r}_{\rm c})/a|$ of the equation of motion, Eq.~(\ref{Eq-of-mot3}),
                vs. radial distance in units of the critical radius ${\hat r}_{\rm c}$, 
	        for a typical O5--V-star in the centre with
		the line acceleration parameters 
		${\hat g_{0}}=17661$, $\gamma=0.4758$, $\delta{}=0.6878$ and ${\hat r_{0}}=1.0016$ in Eq.~(\ref{line-acc-term1}),
		according to $v_{\infty}=3232$~km/s
		(see further stellar parameters in Sect.~\ref{sect-applic-ostar}).
		The horizontal line marks the critical velocity (i.e. sound speed $v_{\rm c}=a$).
                Solution 1 is the unique trans-sonic stellar wind solution through the critical point at ${\hat r}_{\rm c}={\hat r}_{\rm s}=1.0110$
		and ${\hat v}({\hat r}_{\rm c})=1.0$.
                For the description of the different solutions of type 2--6,
                see the discussion in Sects.~\ref{sect-crit-point-sol} and~\ref{sect-sol-eq-mot}.}
\label{sol-topology}
\end{figure}

     The equation of motion (Eqs.~\ref{Eq-of-mot2} and~\ref{Eq-of-mot3})
     yields several families of solutions that have quite different
     mathematical behaviour and physical significance (cf.~Fig.~\ref{sol-topology}). 
     
     The left hand side of Eq.~(\ref{Eq-of-mot3}) vanishes for $(\ud{\hat v}/\ud{\hat r}\neq 0)_{{\hat r_{c}}}$
     at the critical radius ${\hat r_{\rm c}}$, where  
          \begin{equation} \label{crit-velo}   
                   {\hat v}\,({\hat r}_{c}) \equiv {\hat v}\,({\hat r}_{s}) = 1 \, .
          \end{equation} 
     That is, the critical point velocity here is equal to the isothermal sound speed ${\hat v}=1$,
     and the critical radius is just the sonic radius
     \begin{displaymath}
         {\hat r_{\rm c}} \equiv {\hat r_{\rm s}} \, ,
     \end{displaymath}
     as is also the case for thermal winds or mass accretion events (where ${\hat g_{\rm rad}^{\rm line}}\approx{}0$).
     
     We are now interested in under what conditions one can obtain a continuous and smooth trans-sonic flow
     through the critical point ${\hat r}_{\rm c}$ of Eq.~(\ref{Eq-of-mot3}).
     For the case of a stellar wind, this means how to obtain a smooth transition from subsonic and subcritical
     flow \mbox{(${\hat v}<{\hat v}_{\rm c}=1$)} at small 
     \mbox{${\hat r}<{\hat r}_{\rm c}$}
     to supercritical and supersonic flow \mbox{(${\hat v}>{\hat v}_{\rm c}$)} at large
     \mbox{${\hat r}>{\hat r}_{\rm c}$}, 
     when this critical solution has a finite positive slope \mbox{($\ud{\hat v}/\ud{\hat r})>0$}
     at \mbox{${\hat r}={\hat r}_{c}$} (cf.~solid curve 1 in Fig.~\ref{sol-topology})?\footnote{We furthermore assume 
                                                               that both ${\hat v}$ and \mbox{($\ud{\hat v}/\ud{\hat r}$)} are
                                                               {\em everywhere\/} single-valued and continuous.}
     Then, it is evident from the left hand side of Eq.~(\ref{Eq-of-mot3}) that one can obtain such a trans-sonic 
     wind,
     if the right hand side (1)~vanishes at the critical radius ${\hat r}_{\rm c}$,
     (2)~is negative for \mbox{${\hat r}<{\hat r}_{\rm c}$}, and 
     (3)~is positive for \mbox{${\hat r}>{\hat r}_{\rm c}$}.  
 
     The opposite situation occurs for the case of mass accretion in e.g. a collapsing cloud.
     If \mbox{($\ud{\hat v}/\ud{\hat r})_{\hat r_{c}}<0$}, we obtain the second unique trans-sonic and critical solution
     in which ${\hat v}\,({\hat r})$ is monotonically decreasing from supersonic speeds 
     for \mbox{${\hat r}<{\hat r}_{s}$}, e.g. near the protostar, to subsonic speeds for
     \mbox{${\hat r}>{\hat r}_{s}$} at the outer edge of the cloud (see also the second solid line 2, in Fig.~\ref{sol-topology},
     for the case of a corresponding accretion flow with a star as the central object).
     
     Here we are mainly interested in the critical {\em wind} solution
     of Eq.~(\ref{Eq-of-mot3}).
     The right hand side of Eq.~(\ref{Eq-of-mot3}) vanishes at 
     the critical radius ${\hat r}_{c}$ that solves the equation
     \begin{equation}  \label{crit-rad-eq}   
    2\, {\hat r}^{\delta\,\left(1+\gamma\right)} - {\hat v}_{\rm crit}^{2}\, {\hat r}^{\delta\,\left(1+\gamma\right) -1} 
         + {\hat g_{0}}\,\left( {\hat r}^{\delta} - {\hat r_{0}}\right)^{\gamma} = 0\, .
     \end{equation} 
   Therefore, the critical radius has to be determined numerically by means of the above equation and
   the line term pa\-ra\-me\-ters ${\hat g_{0}}$, $\gamma$, $\delta$ and ${\hat r_{0}}$.
   However, if one assumes values of $\gamma$ and $\delta$ close to $1$, one can provide a good approximation (i.e. analytical solution of
   Eq.~(\ref{crit-rad-eq})) for the critical radius
     \begin{equation}  \label{approx-crit-rad}   
      {\hat r}_{\rm c} \approx \frac{\left( {\hat v}_{\rm crit}^{2} - {\hat g_{0}}\right) 
      + \sqrt{\left( {\hat v}_{\rm crit}^{2} - {\hat g_{0}}\right){}^2
      + 8\, {\hat g_{0}}\, {\hat r_{0}} }}{4}    \, .
     \end{equation} 
   For the special case of a thermal wind, where ${\hat g_{0}}$ can be set to zero, we obtain
     \begin{equation} \label{rc-thermalwind}
      {\hat r}_{\rm c} = {\hat r}_{\rm s} = \frac{{\hat v}_{\rm crit}^{2}}{2} = \frac{G\, M \left( 1 - \Gamma \right)}{2 a R} \, .
     \end{equation}
   In some special cases of mass accretion (e.g. a collapsing cloud), one can even set $\Gamma{}=0$.

    \subsubsection{Solving the equation of motion}
   
    The Equation of motion~(\ref{Eq-of-mot3}) can be solved by first integrating the lefthand side over
    ${\hat v}$, and then integrating the righthand side over ${\hat r}$, separately,
    which yields   
         \begin{equation} \label{derive-v-sol-1} 
     {\hat v}^{2} - \ln{\hat v}^{2} 
              = 2\frac{{\hat v}_{\rm crit}^{2}}{\hat r} + 4\, \ln{\hat r}
              + \frac{2}{\hat r_{0}}\,\frac{\hat g_{0}}{\delta\,\left(1+\gamma\right)}
	               \left( 1 - \frac{\hat r_{0}}{{\hat r}^{\delta}}\right)^{1 + \gamma}\!\!\!  + C \, ,
         \end{equation} 
    with the right hand side of Eq.~(\ref{derive-v-sol-1}) denoted as the function
     \begin{equation} \label{def-f-func}
     f\,({\hat r}; {\hat r}', {\hat v}')\! := 2\frac{{\hat v}_{\rm crit}^{2}}{\hat r} + 4 \ln{\hat r}
      + \frac{2}{\hat r_{0}} \frac{\hat g_{0}}{\delta\left(1+\gamma\right)} 
           \left( 1 - \frac{\hat r_{0}}{{\hat r}^{\delta}}\right)^{1 + \gamma}\!\!\! 
           + C({\hat r}',{\hat v}')
     \end{equation}
     with the constant of integration $C$ that is determined by the boundary condition of the radial velocity
    ${\hat v}'$ at a given radius ${\hat r}'$.  
       
     From Eq.~(\ref{derive-v-sol-1}), we can
     determine $C\,({\hat r}',{\hat v}')$ for the solution that passes through the particular point $({\hat r}',{\hat v}')$,
     \begin{equation}
       C\,({\hat r}',{\hat v}') = {\hat v}'^{2} - \ln{\hat v}'^{2} - 2\frac{{\hat v}_{\rm crit}^{2}}{\hat r'} 
          - 4\, \ln{\hat r}' - \frac{2}{\hat r_{0}}\,\frac{\hat g_{0}}{\delta\left(1+\gamma\right)} 
           \left( 1 - \frac{\hat r_{0}}{{\hat r}'^{\delta}}\right)^{1 + \gamma}\, .
      \end{equation}
    Therefore the function $f$ in Eq.~(\ref{def-f-func}) becomes
         \begin{eqnarray}  \label{func-f}
     f\,({\hat r}; {\hat r}', {\hat v}') & = & {\hat v}'^{2} - \ln{\hat v}'^{2} 
                                               + 2\,{\hat v}_{\rm crit}^{2} \left( \frac{1}{\hat r} - \frac{1}{\hat r'}\right)  
					       + 4\,\ln\left(\frac{\hat r}{\hat r'}\right)
                                                \nonumber\\
                                      &  & + \frac{2}{\hat r_{0}}\,\frac{\hat g_{0}}{\delta\left(1+\gamma\right)}  
				             \left[ \left( 1 - \frac{\hat r_{0}}{{\hat r}^{\delta}}\right)^{1 + \gamma}
					                - \left( 1 - \frac{\hat r_{0}}{{\hat r}'^{\delta}}\right)^{1 + \gamma} \right] 
					      \, .    
        \end{eqnarray} 
       And from this, Eq.~(\ref{derive-v-sol-1}) now reads 
     \begin{displaymath}
       {\hat v}^{2} - \ln{\hat v}^{2} = f\,({\hat r}; {\hat r}', {\hat v}')
     \end{displaymath}
     or, equivalently,
       \begin{equation}  \label{derive-v-sol-2}
           -{\hat v}^{2}\, {\rm e}^{-{\hat v}^{2}} = - {\rm e}^{- f\,({\hat r}; {\hat r}', {\hat v}')}\, ,   
       \end{equation} 
     which is solved explicitly and fully analytically in terms of the Lambert W function.

  \subsubsection{The Lambert W function} \label{sec-Lamb-W}
    
    The W function is defined to be the implicit function satisfying
         \begin{equation}  \label{Defgl-W}
             {\rm W} (z)\, {\rm e}^{{\rm W} (z)} = z
         \end{equation}
     (cf.~Corless et al. \cite{corless:et_al:93}, \cite{corless:et_al:96}).   

\begin{figure}
\vspace{-1.0cm}
\centerline{\epsfxsize=9cm \epsfbox{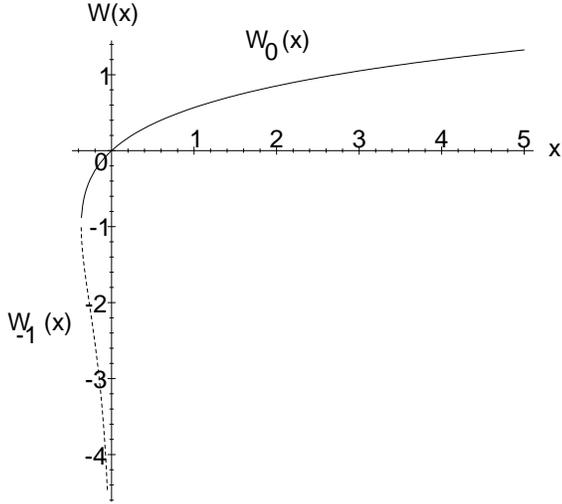}}
\vspace{-1.5cm}
\caption{The two real branches of the Lambert W function: ${\rm W}_{0}(x)$ (full curve) and ${\rm W}_{-1}(x)$ (broken curve).}
\label{wi96f_4}
\end{figure}

     As the equation \mbox{$y(x) \exp y(x)=x$} has an infinite number of solutions $y(x)$ for each non-zero value of
      $x$, W has an infinite number of branches. 
      We are only interested in the physically relevant case, 
      where $x$ is real and \mbox{$-1/e\le{}x<0$}.
      Then, there are two possible real values of W($x$) 
      (see Fig.~\ref{wi96f_4}). 
      As one can already see from the defining equation of W, the W function vanishes at $x=0$,
      is negative for $x<0$, positive for positive values of $x$, and must be $-1$ at the point $x=-1/e$.       
      The branch satisfying \mbox{$-1\le{}{\rm W}(x)$} for $x$ in the range of \mbox{$[-1/e, \infty)$} 
      is denoted by \mbox{${\rm W}_{0}$($x$)} or just W($x$),
      and the branch satisfying \mbox{${\rm W}(x)\leq{}-1$} over the interval \mbox{$[-1/e, 0)$} 
      by \mbox{${\rm W}_{-1}$($x$)}.
      The branches \mbox{${\rm W}_{0}$($x$)} and \mbox{${\rm W}_{-1}$($x$)} share
      the branch point at $x=-1/e$.
      \mbox{${\rm W}_{0}$($x$)} is referred to as the {\em principal branch\/} of the W function, which is the only
      branch that is analytic at 0. 
        The other remaining non-principal branches of W all have a branch point at 0,
      and they are denoted by \mbox{${\rm W}_{k}$($x$)}, where $k$ is a non-zero integer.

    \subsubsection{The solution(s) of the equation of motion}  \label{sect-sol-eq-mot}
    
     It is now possible to provide an explicit analytical expression for the
     solution ${\hat v}$ of the equation of motion~(\ref{Eq-of-mot3}),
     or Eq.~(\ref{derive-v-sol-2}),
     by means of the W function.
     If we compare Eq.~(\ref{derive-v-sol-2}) with
    the defining equation of the Lambert W function (Eq.~\ref{Defgl-W}),
         \begin{displaymath}  
             {\rm W}_{k} (x)\, {\rm e}^{{\rm W}_{k} (x)} = x \, ,
         \end{displaymath}
    we find that
      \begin{displaymath}
     -{\hat v}^{2} = {\rm W}_{k} (x)
      \end{displaymath}
    or
    \begin{equation} \label{vr-gensol-exp1} 
       {\hat v} = \pm \sqrt{- {\rm W}_{k} (x)}
     \end{equation}
   is the {\em general solution} of the equation of motion that passes through the point $({\hat r}',{\hat v}')$,
   with the argument function of the W function
     \begin{equation} \label{arg-func-x} 
       x\,({\hat r}) = - {\rm e}^{- f\,({\hat r}; {\hat r}', {\hat v}')}\, .
     \end{equation}
   Since the argument of the W function in Eq.~(\ref{vr-gensol-exp1}) is always real and negative,
   it is guaranteed that the argument of the square root
   never becomes negative, and hence the solution is always real (see Fig.~\ref{wi96f_4}).  

   Inserting $f\,({\hat r}; {\hat r}', {\hat v}')$ from Eq.~(\ref{func-f}) into Eq.~(\ref{arg-func-x}) yields
     \begin{eqnarray} \label{x-gensol}
     x\,({\hat r};{\hat r}', {\hat v}')\!\! & = & \!\! - \left( \frac{\hat r'}{\hat r} \right)^4 \, 
                          {\hat v}'^{2}\, \exp \left[ 
		       -2\, {\hat v}_{\rm crit}^{2}\, \left( \frac{1}{\hat r} - \frac{1}{\hat r'} \right) \right.
		       \nonumber\\
                       &  &  \left. \!\!\!\! -\frac{2}{\hat r_{0}} \frac{\hat g_{0}}{\delta\left(1+\gamma\right)} \,
                     \left(  \left( 1 - \frac{\hat r_{0}}{{\hat r}^{\delta}}\right)^{1+\gamma} \!\!\! - 
		        \left( 1 - \frac{\hat r_{0}}{{\hat r}'^{\delta}}\right)^{1+\gamma}\right) - {\hat v}'^{2}  
		       \right] \, ,
     \end{eqnarray}
    the general expression for the argument function $x$ depending on the parameters $({\hat r}',{\hat v}')$.
    
    Thus, for the {\em trans-sonic} case of a {\em stellar wind} or general {\em accretion flow}, 
    where ${\hat r}'={\hat r}_{\rm c}\equiv{}{\hat r}_{\rm s}$
    and ${\hat v}'={\hat v_{\rm c}}\equiv{}1$,
    the analytical solution is 
     \begin{equation} \label{vr-windsol} 
       {\hat v}\,({\hat r}) = \pm \sqrt{- {\rm W}_{k} (x\,({\hat r}))} 
     \end{equation}
    with
      \begin{eqnarray} \label{x-windsol} 
     x\,({\hat r}) & = &  - \left( \frac{\hat r_{\rm c}}{\hat r} \right)^4 \, 
                           \exp \left[ 
		       -2\, {\hat v}_{\rm crit}^{2}\, \left( \frac{1}{\hat r} - \frac{1}{\hat r_{\rm c}} \right) \right.
		       \nonumber\\
		   &  &  \left. -\frac{2}{\hat r_{0}} \frac{\hat g_{0}}{\delta\left(1+\gamma\right)} \,
                       \left(  \left( 1 - \frac{\hat r_{0}}{{\hat r}^{\delta}}\right)^{1+\gamma} 
		        - \left( 1 - \frac{\hat r_{0}}{{\hat r_{\rm c}}^{\delta}}\right)^{1+\gamma} \right) - 1  
		       \right] \, .
     \end{eqnarray}
     From this equation, one can easily derive the solution 
     for the simpler cases of a {\em thermal wind} (e.g. the solar wind),
     or for a particular {\em accretion flow}, as e.g. a
     collapsing molecular cloud, 
     where ${\hat g_{0}}$ can be set to zero and ${\hat v}_{\rm crit}^{2}=2\,{\hat r}_{\rm c}$ (cf.~Eq.~\ref{rc-thermalwind}): 
     \begin{equation} \label{vr-thermwind-cloud-sol} 
       {\hat v}\,({\hat r}) = \pm \sqrt{- {\rm W}_{k} \left( - \left( \frac{\hat r_{\rm c}}{\hat r} \right)^4
        \exp\left( 3 - 4\, \frac{\hat r_{\rm c}}{\hat r}\right) \, .
	\right)} 
     \end{equation}
     Here, the minus sign (in front of the square root) refers to the collapsing system. 
     
     We are only interested in the possible two real values of W($x$), the \mbox{$k=0, -1$}--branches
      in Eq.~(\ref{vr-gensol-exp1}) or Eq.~(\ref{vr-windsol}), where $x$ is real and \mbox{$-1/e\le{}x<0$}.
      The branch point at $x=-1/e$, where these two branches meet, 
      corresponds to the critical point ${\hat r}_{\rm c}$, 
      where the velocity in Eq.~(\ref{vr-windsol}) becomes \mbox{${\hat v}={\hat v}({\hat r_{\rm c}})$}$\equiv{}1$.      
      Depending on which branch of W one is approaching this point $x=-1/e$, one obtains a different shape of the 
      ${\hat v} ({\hat r})$--curve, i.e. a stellar wind or a collapsing system.
  
\begin{figure}
\centerline{\hspace{0cm} \epsfxsize=9cm \epsfbox{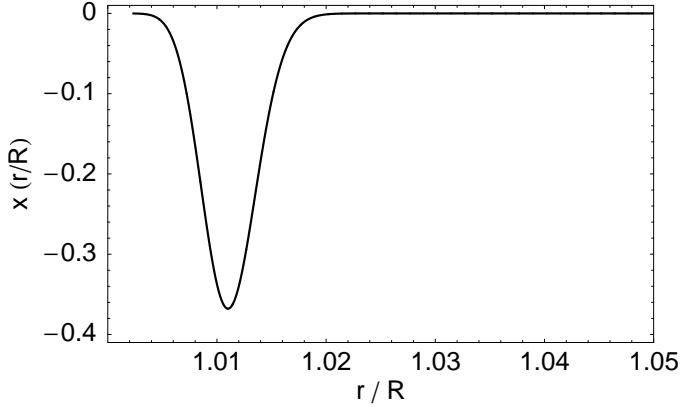}}
\caption{The argument function $x({\hat r})$ of the W function, Eq.~(\ref{x-windsol}), 
                in the trans-sonic velocity law, Eq.~(\ref{vr-windsol}), vs.~radial distance ${\hat r}=r/R$
                for the stellar wind from a typical O5--V-star
		with
		the line acceleration parameters 
		${\hat g_{0}}=17661$, $\gamma=0.4758$, $\delta{}=0.6878$ and ${\hat r_{0}}=1.0016$ in Eq.~(\ref{line-acc-term1}),
		according to $v_{\infty}=3232$~km/s
		(see further stellar parameters in Sect.~\ref{sect-applic-ostar}).
		The minimum of this function is at the critical radius ${\hat r}_{\rm c}={\hat r}_{\rm s}=1.0110$
		where $x({\hat r}_{\rm c})=-1/$e.}
\label{arg-func}
\end{figure}

      However, to determine which of the two branches to choose 
      at a certain range of radius between \mbox{$[1, {\hat r}_{\rm c}]$} and \mbox{$[{\hat r}_{\rm c}, \infty )$}
      so as to guarantee a continuous, monotonically increasing,
      and smooth trans-sonic flow (as in the case of a wind),
      one still needs to examine the behaviour of the argument function $x({\hat r})$ of $W$, in Eq.~(\ref{x-windsol}),
      with radius ${\hat r}$ (see Fig.~\ref{arg-func}).
      
      The argument func\-tion \mbox{$x({\hat r})$}
      decreases monotonically from the stellar radius ${\hat r}=1$ (with a value of nearly zero) to its minimum at 
      ${\hat r}={\hat r}_{\rm c}$ with $x=-1/e$         
      to afterwards increase monotonically again.
 
       {\sc Stellar wind:}~~By            
       choosing first the real principal branch \mbox{${\rm W}_{0}$($x$)} of the Lambert function 
       that is followed
       in the negative direction until the critical point \mbox{$x({\hat r}_{\rm c})=-1/e$} and \mbox{${\rm W}_{0}$($x$)$=-1$}
       (i.e.~\mbox{${\hat v} ({\hat r}_{\rm c})$)}
       for rising radii from near the star surface ${\hat r}\approx{}1$, and hence
       coming from \mbox{$x({\hat r})\lesssim{}0$} and 
       \mbox{${\rm W}_{0}$($x$)$\lesssim{}0$},
       yields the desired critical wind solution.
       This choice causes the strict monotonic increase of the velocity \mbox{${\hat v}({\hat r})$}
       from nearly zero until the sonic point (in the subsonic region).
       To ensure that the solution passes smoothly through this critical point, i.e.~that the derivative
        \mbox{${\hat v'}({\hat r}_{\rm c})$} is continuous,
       one has to change there to the \mbox{${\rm W}_{-1}$($x$)}--branch.       
       From there onwards, the real part of this branch is followed further in the negative direction
       \mbox{${\rm W}_{-1}$($x$)$\rightarrow{}-\infty$} for larger radii and therefore with
       decreasing $|x({\hat r})|\rightarrow{}0$, again.  
       From this, one obtains the desired continued monotonic slope of \mbox{${\hat v}({\hat r})$} in the supersonic
       region for ${\hat r}>{\hat r_{\rm c}}(={\hat r_{\rm s}})$. 

      {\sc Accretion solution:}~~In this case, the unique critical and trans-sonic solution
       \mbox{${|\hat v}({\hat r})|$} is strictly monotonically
       increasing from small values in the subsonic region far away from the central object
       (for ${\hat r}>{\hat r_{\rm s}}$ and $|{\hat v}({\hat r})|\ll{}1$),
       into the supersonic region (for ${\hat r}<{\hat r_{\rm s}}$ and $|{\hat v}({\hat r})|>1$)
       with declining radius ${\hat r}$. Here
       the slope ${\hat v'}$ is negative and continuous at the sonic (critical) radius ${\hat r_{\rm s}}$
       (as a necessary boundary condition).
       This case can be achieved when we again first take  
       the \mbox{${\rm W}_{0}$($x$)}--branch for \mbox{$0\ge{}x({\hat r})\ge{}-1/e$}
       (in the subsonic region)
       and then change over at the sonic point to
       the \mbox{${\rm W}_{-1}$($x$)}--branch for \mbox{$-1/e\le{}x({\hat r})<0$}; now however with 
       decreasing radius ${\hat r}\rightarrow{}1$. 
       Here one is again wandering in the same negative direction through the possible two
       real values of W($x$) in the range of \mbox{$-1/e\le{}x<0$}.
       We are approaching the critical point \mbox{$x({\hat r}_{\rm s})=-1/e$}
       (where the argument function $x({\hat r})$ has its minimum, cf. Fig.~\ref{arg-func})
       with declining radii ${\hat r}>{\hat r_{\rm s}}$ from the right side and
       from $x({\hat r})\lesssim{}0$,
       in order to change the branch at this point, where 
       $x({\hat r})$ afterwards increases again strictly monotonically (with smaller radii) until almost
       $x({\hat r}_{\star})\lesssim{}0$ near the central object.
          
    To summarise, the velocity in the radial direction (i.e.~the exact analytical and spherically
    trans-sonic solution of our equation of motion) can be described,       
    \begin{itemize}
    \item[(a)] for the case of a {\em stellar wind},
        by Eq.~(\ref{vr-windsol}) (and the positive sign in front of the root) 
	and the argument function in Eq.~(\ref{x-windsol}),
	choosing the branch
	\begin{equation} \label{k-branch-wind}
 	 k =
          \left\{  \begin{array}{ccc}
           \,\,\, 0 & \mbox{for} & 1 \leq {\hat r} \leq {\hat r_{\rm c}} \\
           -1 & \mbox{for} & {\hat r} > {\hat r_{\rm c}} 
                       \end{array}
          \right. 
	\end{equation}
	of the W function at a certain radius ${\hat r}$,
	and 
    \item[(b)] in case of a {\em general accretion flux}, as well,
        by Eq.~(\ref{vr-windsol}) (but now with the negative sign in front of the root) 
	and the argument function in Eq.~(\ref{x-windsol}),
	choosing the branch	   
  	\begin{equation} \label{k-branch-accretion}
 	 k =
          \left\{  \begin{array}{ccc}
           -1 & \mbox{for} & 1 \leq {\hat r} \leq {\hat r_{\rm c}} \\
          \,\,\, 0 & \mbox{for} & {\hat r} > {\hat r_{\rm c}} 
                       \end{array}
          \right. 
	\end{equation}   
         depending on the location ${\hat r}$, and
    \item[(c)] in the special cases of a {\em thermal wind or} collapsing system like a
        {\em collapsing protostellar cloud},
	by Eq.~(\ref{vr-thermwind-cloud-sol}),
	choosing the same branches as mentioned above in case (a), or (b) respectively,
	and the appropriate sign (in front of the root).    
     \end{itemize}
    In addition to these two critical solutions (type 1 and 2, cf.~numbering in Fig.~\ref{sol-topology}),
     already discussed, 
     that pass through the critical point (i.e. sonic point), all the other four types of solutions were obtained from our {\em general
     velocity law}, Eq.~(\ref{vr-gensol-exp1}) with Eq.~(\ref{x-gensol}),
     choosing the following branches of the W function and 
     values of \mbox{(${\hat r}',{\hat v}'$)}, for the point we demand the solution to go through:
     \begin{itemize}
     \item Type-3: {\em Subsonic (subcritical) solutions}
                    \begin{displaymath}
                     k=0\, , \quad  {\hat r}'= {\hat r}_{\rm c} , \quad {\hat v}'<1
                    \end{displaymath}
     \item Type-4: {\em Supersonic (supercritical) solutions}
                    \begin{displaymath}
                     k=-1\, , \quad  {\hat r}'= {\hat r}_{\rm c} \, , \quad {\hat v}'>1
                    \end{displaymath}
     \item Type-5 \& 6: {\em Double-valued solutions} 
                     \begin{displaymath}
                     k=0 \quad \mbox{and} \quad k=-1 \quad \mbox{for}
                    \end{displaymath}
                     \begin{displaymath}
                     {\hat r}'\, , {\hat v}' = \mbox{arbitrarily,\, where} \quad {\hat r}'\neq{}{\hat r}_{\rm c} \, , \quad {\hat v}'\neq{}1\, .
                    \end{displaymath}
     \end{itemize}
      Type-3 solutions are everywhere subsonic
      (choosing only the principal branch, $k=0$, of the W function).
      Those of type 4 are everywhere supersonic
      (selecting only the $k=-1$-branch in the velocity law),
      and those of type 5 and 6 are double-valued,
      composed of both the $k=0$ and $k=-1$-branch, below and above the sonic line, respectively.
       In this connection, the two sub- and supersonic pairs of curves of this last mentioned types,
      subdivided into \mbox{(5a, 6a)} and \mbox{(5b, 6b)} in Fig.~\ref{sol-topology}, belong together.
      They are fixed, not only by the same chosen branch of W in Eq.~(\ref{vr-gensol-exp1}),
      but also by the same selected parameters for the solution through the identical given point 
      \mbox{(${\hat r}',{\hat v}'$)}.
 
     From our analytical wind solution,
     we find that ${\hat v}$ increases without limit as ${\hat r}\rightarrow{}\infty$.
     This unphysical increase in ${\hat v}$ at large ${\hat r}$ is an artificial consequence
     of our assumption (1) that the stellar wind is isothermal with a temperature, $T_{\rm eff}$, 
     of the central stellar photosphere.
     This assumption is valid close to the photosphere, but may not hold at large distances, where 
     the gas expansion will ultimately force the wind to cool down.
     By insisting on keeping the wind temperature $T$ fixed, we introduce an undesirable energy source
     to do work on the gas and accelerate it without limit.
     In the case of the solar wind, this classical problem   
     was overcome in the work of Parker (\cite{parker}) (see also the review of Mihalas \cite{mihalas})
     by assuming the corona near the sun to be isothermal, yet allowing
     the wind to expand adiabatically at larger radii.
     
     We overcome this problem by deriving an analytical expression
     for the wind solution in the supersonic approximation, in the following Subsect.~\ref{sect-approxsol-eq-mot},
     by neglecting the pressure term in the equation of motion
     which introduces a terminal velocity $v_{\infty}$, as well as a relationship between $v_{\infty}$
     and our wind (and line acceleration) parameters ${\hat g}_{0}$, $\gamma$, $\delta$ and ${\hat r_{0}}$.
      
   \begin{figure}
    \centerline{\epsfxsize=9cm \epsfbox{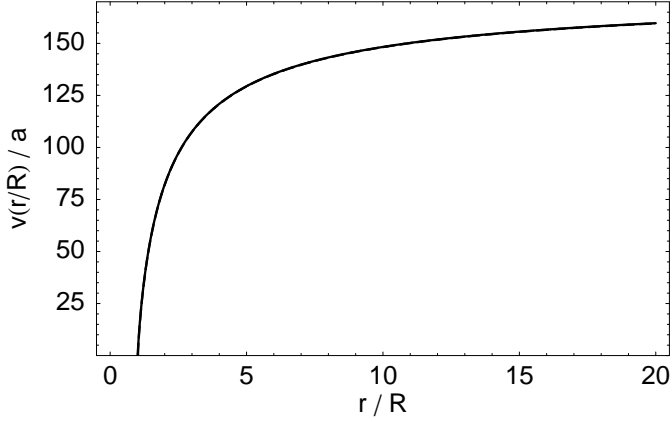}}
    \caption{Comparison of the approximated wind solution (invisible dashed line, overlaid by the solid line) with the exact solution
             ${\hat v}\,({\hat r})$ (in units of sound speed) vs. radial distance ${\hat r}$ (in units of $R=11.757\,R_{\sun}$)
	     for the wind from a typical O5--V-star
             with the line acceleration parameters 
	     ${\hat g_{0}}=17661$, $\gamma=0.4758$, $\delta{}=0.6878$ and ${\hat r_{0}}=1.0016$ in Eq.~(\ref{line-acc-term1}),
             according to $v_{\infty}=3232$~km/s
	     (see stellar parameters in Sect.~\ref{sect-applic-ostar}).
             The approximated curve was obtained from Eq.~(\ref{vwind-sol-approx}),
	     where the exact solution was obtained from Eq.~(\ref{vr-windsol}), together with Eq.~(\ref{x-windsol}) and
	     Eq.~(\ref{k-branch-wind}),
	     with a critical radius of ${\hat r}_{\rm c}=1.0110$.
	     For a more detailed comparison see Fig.~\ref{pic-v-approx-ex-beta}.}
    \label{pic-v-approx-ex}
    \end{figure}  
   
   \begin{figure}
    \centerline{\epsfxsize=9cm \epsfbox{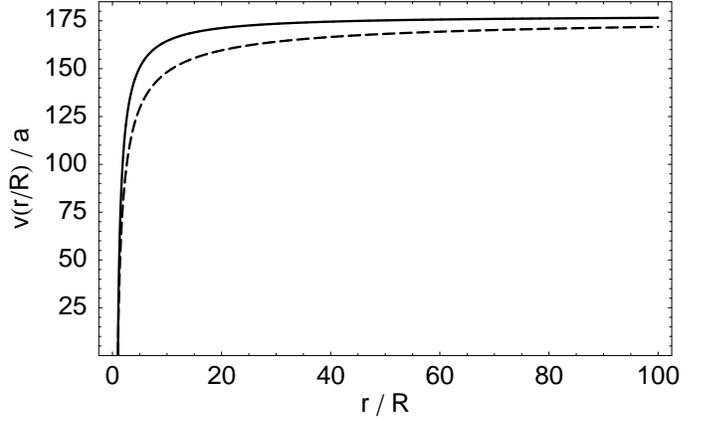}}
    \caption{Comparison of the approximated wind solution (dashed line) with the $\beta$ wind velocity law (solid line)
             ${\hat v}\,({\hat r})$ (in units of sound speed a) over a large radial distance ${\hat r}$ (in units of $R=11.757\,R_{\sun}$)
	     for the wind from a typical O5--V-star.
	     The approximated wind curve was obtained 
	     with the same parameters as in Fig.~\ref{pic-v-approx-ex},    
	     where the $\beta$ wind velocity curve was retrieved from Eq.~(\ref{beta-law})
	     with an exponent of $\beta=0.7379$,
	     a terminal velocity of $(v_{\infty}/a)=177.9$ (according to $v_{\infty}=3232$ km/s) 
	     and a parameter value of ${\hat r_{0}'}=1.0095$ (used in {\sc ISA-Wind}).
	     Both curves approach the same velocity limit $v_{\infty}$ as ${\hat r}$ goes to infinity.
	     }
    \label{pic-v-approx-beta}
    \end{figure}     
     
\begin{figure}
\centerline{\hbox{\hspace{0cm}\epsfxsize=9cm \epsfbox{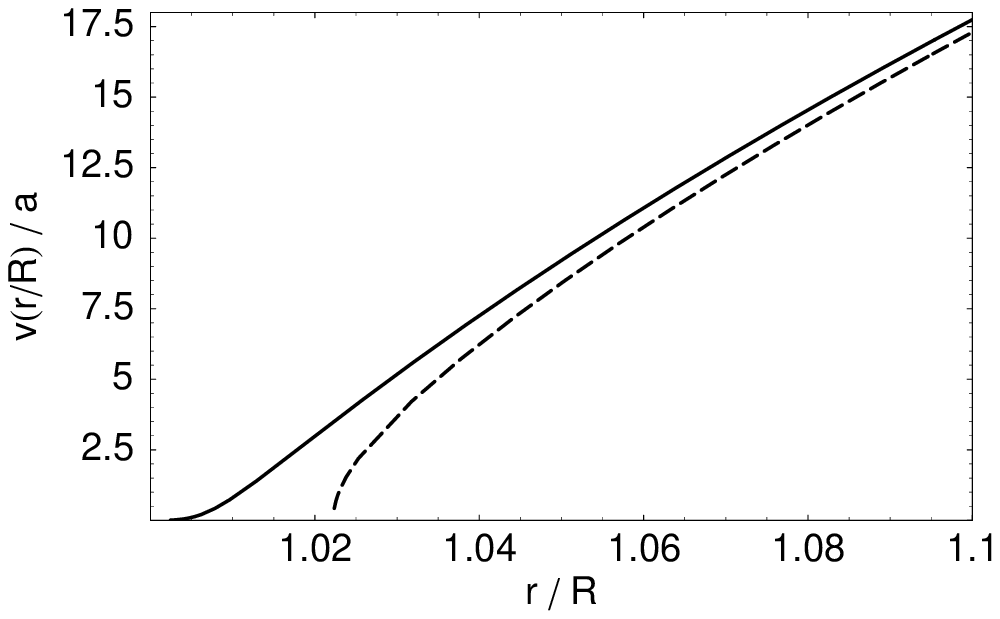}}}
\centerline{\hbox{\hspace{-0.2cm} \epsfxsize=9cm \epsfbox{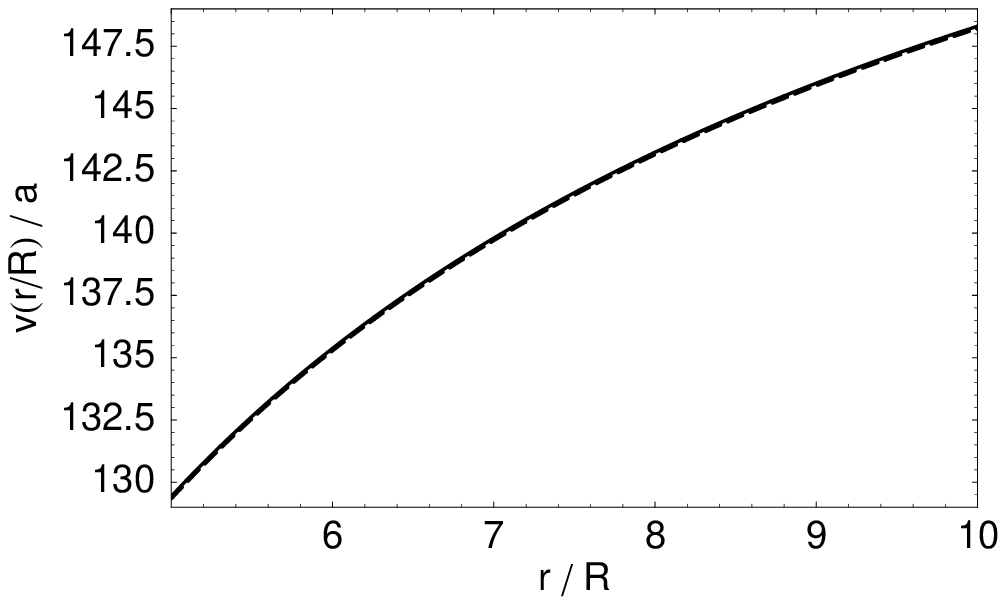}}}
\centerline{\hbox{\hspace{0cm}\epsfxsize=9cm \epsfbox{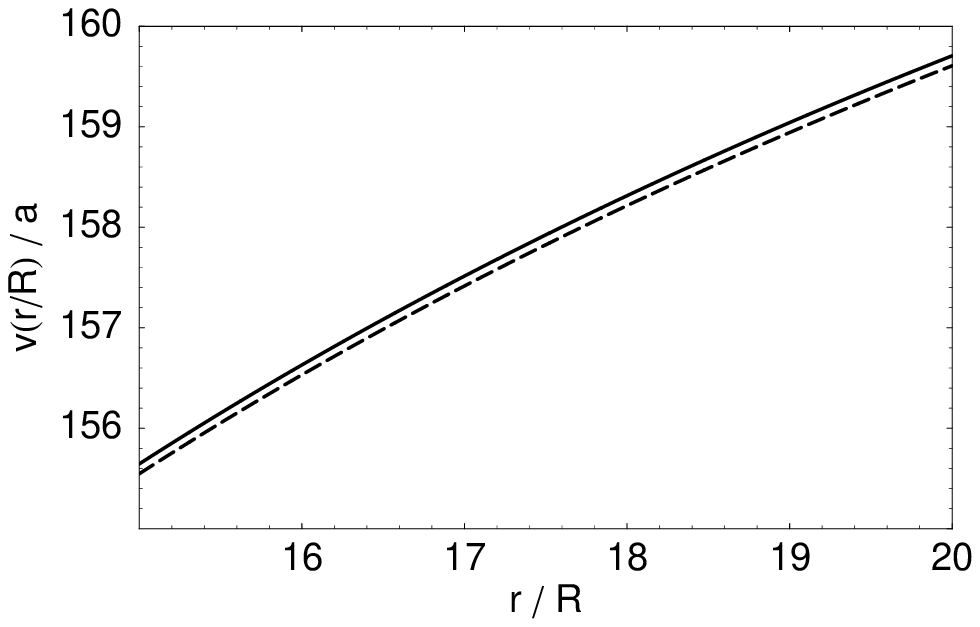}}}
\caption{Model results for the wind velocity
             ${\hat v}\,({\hat r})$ (in units of sound speed a) vs. radial distance ${\hat r}$ (in units of $R=11.757\,R_{\sun}$)
	     from a typical O5--V-star
             with the line acceleration parameters 
	     ${\hat g_{0}}=17661$, $\gamma=0.4758$, $\delta{}=0.6878$ and ${\hat r_{0}}=1.0016$
	     (cf. Eq.~\ref{line-acc-term1}),
	     and a critical radius of ${\hat r}_{\rm c}={\hat r}_{\rm s}=1.0110$
             (see stellar parameters in Sect.~\ref{sect-applic-ostar}). 
	     The approximated (dashed line) and the exact wind solution (solid line) are compared with each other
	     for three different and magnified radial ranges:
	     Very close to the central star between ${\hat r}=[1.0,1.1]$ (see upper diagram), 
	     at intermediate distance of ${\hat r}=[5.0,10.0]$ (see middle diagram)
	     and far away between ${\hat r}=[15.0,20.0]$ (see lower diagram).}
\label{pic-v-approx-ex-beta}
\end{figure}

     At the end of the iteration process (described in Sect.~\ref{sect-it-method}), 
     we obtain the converged values of the previously mentioned wind model parameters, together with the real position of the critical radius
     ${\hat r}_{\rm c}$, to
     be able to evaluate the final
     wind solution with Eqs.~(\ref{vr-windsol}) and~(\ref{x-windsol}).
     This way, the complete (exact) wind velocity curve is produced (in Fig.~\ref{pic-v-approx-ex}) to be
     compared with the approximated wind solution over a wide wind range from the stellar photosphere up to a distance of 20 stellar
     radii.
     This comparison (see also the enlarged images of Fig.~\ref{pic-v-approx-ex-beta}) shows a high level of agreement, and only a weak 
     discrepancy between both curves ${\hat v}\,({\hat r})$, in particular for larger radii, where the (even) unlimited function for
     the exact wind solution is close to the 'speed' limited approximated solution curve.
     Through this way of adjustment, the undesirable unlimited mathematical behaviour of our expression for the wind velocity
     has become insignificant over the whole ${\hat r}$ range of interest.
     
     The subsequently derived approximated wind solution is only valid in the supersonic region and is not supposed to be applied to the 
     subsonic region,
     where it even becomes imaginary, particularly in our wind model in the range of $0<{\hat r}\lesssim{}{\hat r}_{\rm s}$,
     see upper diagram of Fig.~\ref{pic-v-approx-ex-beta}.  
     This explains the only noticeable discrepancy between both wind velocity curves in that radial range, from close above the sonic radius down
     to the stellar surface.
   
     The numerically computed model atmospheres 
     that we apply later in our iteration process
     allow a temperature stratification $T\,(r)$ of the wind, i.e. we do not assume 
     a fixed $T$ as in our analytical expressions.

   \subsubsection{Approximated solution of the equation of motion}  \label{sect-approxsol-eq-mot}

     By neglecting the pressure term in the Eq. of motion~(\ref{Eq-of-mot1-nd}), 
     \begin{displaymath}
        \frac{1}{\rho}\, \frac{\ud \rho}{\ud{\hat r}} \approx 0 \, ,
     \end{displaymath}
     which is a good approximation
     for the stellar wind solution in the supersonic region with ${\hat r}>{\hat r}_{\rm c}\equiv{}{\hat r}_{\rm s}$,
     Eq.~(\ref{Eq-of-mot3}) becomes
    \begin{equation}  \label{Eq-of-mot-approx}  
         {\hat v} \, \frac{\ud}{\ud\,{\hat r}}\, {\hat v} =
                -\frac{{\hat v_{\rm crit}}^{2}}{{\hat r}^{2}} 
                + \frac{\hat g_{0}}{\hat r^{1+\delta\,\left( 1+\gamma\right)}}\, \left({\hat r}^{\delta} - {\hat r_{0}}\right)^{\gamma} \, .
     \end{equation}
   This simplified equation of motion can again be solved by first integrating the left hand side over
    ${\hat v}$, and then integrating the right hand side over ${\hat r}$, separately,
    which yields   
         \begin{equation} \label{derive-v-sol-approx} 
     {\hat v}^{2} = 2\frac{{\hat v}_{\rm crit}^{2}}{\hat r}
                  + \frac{2}{\hat r_{0}}\,\frac{\hat g_{0}}{\delta\,\left(1+\gamma\right)}
		   \left( 1 - \frac{\hat r_{0}}{{\hat r}^{\delta}}\right)^{1 + \gamma} + C \, .
         \end{equation} 
   In order to determine the integration constant $C$, we assume a boundary condition
   \mbox{${\hat v}\,({\hat r'}) \approx 0$}
   for the wind velocity at radius ${\hat r'}={\hat r_{0}}^{1 / \delta}$, close to the stellar photosphere, i.e.
    \begin{equation}
    C\,({\hat r}'\!=\!{}{\hat r_{0}^{1 / \delta},{\hat v}'\!=\!{}0) = -2\, \frac{{\hat v}_{\rm crit}^{2}}{{\hat r_{0}}^{1 / \delta}}}\, .
    \end{equation}
    Thus, Eq.~(\ref{derive-v-sol-approx}) now reads 
    \begin{equation}
     {\hat v}^{2} = 2\,\left( {\hat v}_{\rm crit}^{2} \left( \frac{1}{\hat r} - \frac{1}{{\hat r_{0}}^{1 / \delta}} \right) 
                    + \frac{\hat g_{0}}{{\hat r_{0}}\,\delta \left( 1+\gamma\right)} 
		         \left( 1 - \frac{\hat r_{0}}{{\hat r}^{\delta}}\right)^{1 + \gamma} \right) \, ,
    \end{equation}
    resulting in the {\em approximated wind solution}
    \begin{equation} \label{vwind-sol-approx} 
     {\hat v}\,({\hat r}) = \sqrt{ \frac{2}{\hat r_{0}}\,\left( {\hat v}_{\rm crit}^{2} 
                      \left( \frac{\hat r_{0}}{\hat r} - {\hat r_{0}}^{1 - 1 / \delta} \right) 
                    + \frac{\hat g_{0}}{\delta \left( 1+\gamma\right)} \left( 1 - \frac{\hat r_{0}}{{\hat r}^{\delta}}\right)^{1 + \gamma}\right)
                      }\, ,
    \end{equation}
    which can be expressed without the W function.

    \subsubsection{Comparison with the $\beta$ velocity law}  \label{sect-comp-beta-law}

   Eq.~(\ref{vwind-sol-approx}) yields a terminal velocity ${\hat v}_{\infty}$ (as ${\hat r}\rightarrow{}\infty$) of
     \begin{equation} \label{vinf-law} 
      {\hat v}_{\infty} = \sqrt{ \frac{2}{\hat r_{0}}\,\left( \frac{{\hat g_{0}}}{\delta \left(1 + \gamma\right)}
                                   -  {\hat v}_{\rm crit}^{2}\, {\hat r_{0}}^{1 - 1 / \delta}  \right)} \, ,
     \end{equation}
    which is now comparable to the ${\hat v}_{\infty}$ parameter in a (so-called) 
    $\beta$ velocity law (cf. Castor \& Lamers 1979, CAK)
    \begin{equation} \label{beta-law} 
     {\hat v}\,({\hat r}) = {\hat v}_{\infty} \,  \left( 1 - \frac{\hat r_{0}'}{\hat r}\right)^{\beta} \, .
    \end{equation}
    To be able to compare the $\gamma$ (and $\delta$) parameter in our wind law with the exponent in the
    $\beta$ law (as we use $\beta$ as an input parameter in our model atmosphere calculations),  
   we express our line acceleration parameter ${\hat g_{0}}$ in terms of ${\hat v}_{\infty}$
   by means of Eq.~(\ref{vinf-law}), i.e.
   \begin{equation} \label{g0-vinf-rel} 
     {\hat g_{0}} = \left( \frac{{\hat v}_{\infty}^2}{2} + \frac{{\hat v}_{\rm crit}^{2}}{{\hat r_{0}}^{1/\delta}}\right)\,
                    {\hat r_{0}}\, \delta \left( 1 + \gamma \right)\, ,
   \end{equation}
   and insert it into Eq.~(\ref{vwind-sol-approx}), as to obtain
   \begin{equation} \label{vwind-sol-approx2} 
    {\hat v}\,({\hat r}) =\!\!\! \sqrt{\frac{2}{\hat r_{0}}\, \left( {\hat v}_{\rm crit}^{2} 
                          \left( \frac{\hat r_{0}}{\hat r} - {\hat r_{0}}^{1 - 1/\delta} \right) 
                        + \left( \frac{\hat r_{0}}{2} {\hat v}_{\infty}^2  + {\hat v}_{\rm crit}^{2} {\hat r_{0}}^{1 - 1/\delta} \right)
			 \left( 1 - \frac{\hat r_{0}}{{\hat r}^{\delta}}\right)^{1 + \gamma} \right)} \, ,
   \end{equation}
   which now depends on ${\hat v}_{\infty}$, $\gamma$, $\delta$ and ${\hat r_{0}}$.
   
   Since $\delta$ is of the order of 1, as is ${\hat r_{0}}$,
   one can approximate the expression ${\hat r_{0}}^{1 - 1/ \delta}$ in Eq.~(\ref{vwind-sol-approx2})
   as 1. 
   Furthermore, as our line parameter ${\hat r_{0}}$
   is defined as the parameter for which the line acceleration becomes zero at radius ${\hat r}={\hat r_{0}}^{1 / \delta}$
   (cf.~Eq.~\ref{line-acc-term1}), this radius is very close to the radius ${\hat r_{0}}'$
   in the $\beta$--law (in Eq.~\ref{beta-law}),
   where the wind velocity is assumed to be zero, i.e. ${\hat r_{0}}'\approx{}{\hat r_{0}}$.
   
   Then, we can set the velocity in Eq.~(\ref{beta-law}) equal to our velocity law in Eq.~(\ref{vwind-sol-approx2}),
   to search for a relationship between the parameters $\beta$, $\gamma$ and $\delta$,
   which yields
   \begin{equation}
    \left( 1 - \frac{\hat r_{0}}{\hat r}\right)^{2\beta - 1} \stackrel{!}{\approx}  - b + \left( 1 + b \right)\, 
    \frac{\displaystyle \left( 1 - \frac{\hat r_{0}}{{\hat r}^{\delta}}\right)^{1+\gamma}}
                                     {\displaystyle \!\!\!\!\! \left( 1 - \frac{\hat r_{0}}{\hat r}\right)}
   \end{equation}
   with
   \begin{displaymath}
    b := \frac{2}{\hat r_{0}} \left(\frac{{\hat v}_{\rm crit}}{{\hat v}_{\infty}}\right)^2 \, .
   \end{displaymath}
   
   Herein, for large radii ${\hat r}$ (and especially for small values of $b$, e.g.~$b\lesssim{}0.1$ for an O-V-star),
   the right hand side can be approximated by the last fraction only,
   which leads to
   \begin{equation} \label{beta-gam-del-rel1}
   \left( 1 - \frac{\hat r_{0}}{\hat r}\right)^{2\beta} \stackrel{!}{\approx} \left( 1 - \frac{\hat r_{0}}{{\hat r}^{\delta}}\right)^{1+\gamma}\, ,
   \end{equation}
   or equivalently
   \begin{equation} \label{beta-gam-del-rel}
   \frac{2 \beta}{1+\gamma} \approx \frac{\displaystyle \log \left( 1 - \frac{\hat r_{0}}{{\hat r}^{\delta}}\right)}
                                                             {\displaystyle \log \left( 1 - \frac{\hat r_{0}}{\hat r}\right)} \, . 
   \end{equation}
   Next, the righthand side in Eq.~(\ref{beta-gam-del-rel}) can be approximately set to 1,
   for values of $\delta\longrightarrow{}1$ or generally for smaller distances from the central star
   in the supersonic region as ${\hat r}\longrightarrow{}{\hat r_{0}}^{1 / \delta}\approx{}1$.
   This results the desired relationship between $\beta$ and $\gamma$
   \begin{equation} \label{beta-gam-rel}
    2\, \beta  \approx  1 + \gamma   \, ,
   \end{equation}
   independent of $\delta$, that is most
   valid for the previously mentioned values for $\delta$ at smaller radii ${\hat r}$.
   It also applies at very large distances ${\hat r}\gg{}1$, since then, the numerical values inside the brackets of Eq.~(\ref{beta-gam-del-rel1}) 
   are close to 1 and this equation is fulfilled for any value of the exponents $\beta$ and $\gamma$ in all cases.
   Only for intermediate distances from the star at lower values of $\delta$ (not close to 1),
   the relationship between $\beta$ and $\gamma$ is possibly not well approximated by
   Eq.~(\ref{beta-gam-rel}).
   
   Figure~\ref{pic-v-approx-beta} illustrates this 
   described behaviour by the comparison of the approximated wind solution with the $\beta$ law for our wind model calculations
   from a typical O5--V-star.

   \subsubsection{Fitting formula for the line acceleration}   \label{sect-fit-form-gline}
   
   Thus, we can provide another expression for the line acceleration (in Eq.~\ref{line-acc-term1}),
   now dependent on ${\hat v}_{\infty}$ and on the parameters $\beta$ (or $\gamma$ respectively), $\delta$ and ${\hat r_{0}}$
   (by eliminating ${\hat g_{0}}$ using Eq.~\ref{g0-vinf-rel}):
    \begin{equation} \label{line-acc-term3a}
    {\hat g_{\rm rad}^{\rm line}} ({\hat r}) 
    = \left( \frac{{\hat v}_{\infty}^2}{2} + \frac{{\hat v}_{\rm crit}^{2}}{{\hat r_{0}}^{1/\delta}}\right)\,
                    {\hat r_{0}}\, \delta \left( 1 + \gamma \right)\, \frac{1}{{\hat r}^{1 + \delta}}\,
                    \left( 1 - \frac{\hat r_{0}}{{\hat r}^{\delta}} \right)^{\gamma}   \, ,
    \end{equation}
    or
    \begin{equation} \label{line-acc-term3b}
    {\hat g_{\rm rad}^{\rm line}} ({\hat r}) 
    = \left( {\hat v}_{\infty}^2 + \frac{2\,{\hat v}_{\rm crit}^{2}}{{\hat r_{0}}^{1/\delta}}\right)\,
                    {\hat r_{0}}\, \delta \, \beta \, \frac{1}{{\hat r}^{1 + \delta}}\,
                    \left( 1 - \frac{\hat r_{0}}{{\hat r}^{\delta}} \right)^{2 \beta - 1}   \, .
    \end{equation}
    Both of these non-linear expressions can then be used as fitting formulae and applied to
    the results from a numerical calculation of ${\hat g_{\rm rad}^{\rm line}}\,({\hat r}_{i})$
    for discrete radial grid points ${\hat r}_{i}$,
    in order to determine the line acceleration parameters $\gamma$ (or equivalently $\beta$),
    $\delta$ and ${\hat r_{0}}$,
    and the terminal velocity ${\hat v_{\infty}}$ (cf. Fig.~\ref{pic-gline-fit}).

   \subsubsection{Physical interpretation of the equation of critical radius}  \label{sect-interpret-eq-of-critrad}

   Through the use of the exact wind solution, by using Eq.~(\ref{crit-rad-eq}), valid for the critical radius ${\hat r}_{\rm c}$,
   we can solve for the line acceleration parameter ${\hat g}_{0}$
   and insert it into Eq.~(\ref{vinf-law}) from the approximated wind solution,
   to provide another expression for the terminal velocity
   depending on the location of the critical (i.e. sonic) point
	  \begin{equation} \label{v-inf-rc}
	{\hat v}_{\infty} = \sqrt{\frac{2}{\hat r_{0}} \left( 
         \left( \frac{ {\hat r}_{\rm c}^{\delta} }{ {\hat r}_{\rm c}^{\delta} - {\hat r_{0}}}\right)^{\gamma} 
	   \frac{ {\hat r}_{\rm c}^{\delta - 1} }{\delta \left( 1+\gamma\right)} \left( {\hat v}_{\rm crit}^{2} 
	       - 2\, {\hat r}_{\rm c}\right) - {\hat v}_{\rm crit}^{2} {\hat r_{0}}^{1 - 1/ \delta} \right)} \, .
	  \end{equation}
   
    Or vice versa,
    by Eq.~(\ref{crit-rad-eq}), the location of the critical point
    (through which the exact analytical wind solution of our equation of motion EOM~(\ref{Eq-of-mot3}) passes) 
    is mainly determined, on the one hand, by the given terminal velocity $v_{\infty}$, via the line acceleration parameter
    ${\hat g_{0}}$.
    On the other hand, the position of ${\hat r_{\rm c}}$ must also be dependent on the given minimum velocity $v_{\rm in}$
    at the inner boundary radius $R_{\rm in}$, where the velocity solution passes.
    This inner velocity $v_{\rm in}$ follows indirectly from the other remaining line acceleration or wind parameters
    $\gamma$, $\delta$ (which make up the shape of the velocity curve) and especially ${\hat r_{0}}$
    (where the value of the latter parameter is determined by the radius ${\hat r_{0}}^{1/\delta}$ at which $g_{\rm rad}^{\rm line}$ is zero).
      
    Since the inner boundary condition of the velocity $v_{\rm in}$ is connected to the mass-loss rate $\dot{M}$,
    through the equation of mass continuity by Eq.~(\ref{Mdot}) and the given density at the inner boundary,
    the position of the critical radius 
    is uniquely specified by the values of $v_{\infty}$ and $\dot{M}$.

   \subsubsection{The reasons for deriving the approximated solution}   \label{sect-reason-for-approxsol}  
    
    The reason for the derivation of our \emph{approximated} wind solution (in Sect.~\ref{sect-approxsol-eq-mot}),
    that is only valid in the supersonic approximation,
    is \emph{not} its application to represent the unique wind solution in the whole wind regime
    (i.e. including the subsonic region).  
    The approximated solution is required to be able to calculate the terminal velocity $v_{\infty}$  
    belonging to our exact analytical wind solution described by Eqs.~(\ref{vr-windsol}) and~(\ref{x-windsol})
    and to overcome its artificial and unlimited increase with increasing radius
    by assuming an isothermal wind (as discussed at the end of Sec.~\ref{sect-sol-eq-mot}).
    
    In other words:~the approximated solution is necessary to be able
    (1) to find the relation between $v_{\infty}$ and the line acceleration (or wind) parameters ${\hat g_{0}}$, $\gamma$, $\delta$
         and ${\hat r_{0}}$, and
    (2) to compare the exponents $\gamma$ and $\delta$ to the exponent $\beta$ (in the $\beta$-law). 
    
    This, in turn, is necessary because
    our applied model atmosphere code assumes a $\beta$-velocity law for the whole supersonic wind region
    which is mainly determined by the input parameters $v_{\infty}$ and $\beta$.
    The additionally used code parameter ${\hat r_{0}}'$ (cf. Eq.~\ref{beta-law})
    is related to the velocity $v_{\rm in}$ at the inner boundary radius
    which, again, is directely dependent on the prescribed mass-loss rate $\dot{M}$
    via the continuity Eq.~(\ref{Mdot}) (and the density at the inner boundary given therein).

    \section{Numerical methods}  \label{sect-num-meth}

    \subsection{Computing the radiative acceleration} \label{sect-num-comp-gline}

    The calculation of the radiative acceleration of a stellar wind requires the numerical computation of
    the contribution of a very large number of spectral lines.
    We first calculate the thermal, density and ionisation structure of the wind model by means of the
    non-LTE expanding atmosphere (improved Sobolev approximation) code {\sc ISA-Wind} 
    (de Koter et al.~\cite{dekoter:schmutz}, \cite{dekoter:heap}).
    As a next step, we calculate the radiative acceleration
    as a function of distance by means of a Monte Carlo (MC) code {\sc MC-Wind} 
    (de Koter et al. \cite{dekoter:heap}, Vink et al. \cite{vink:dekoter:lamers}).
    The basic ideas behind this technique were first applied to the study of stellar winds
    from early-type stars by Abbott \& Lucy (\cite{abbott:lucy}), but in our calculation we  
    account for the possibility that the photons
    can be scattered or absorbed and re-emitted (due to real absorption) or eliminated (if they are scattered back into the star).
     
    The radiative transfer in {\sc MC-Wind} involves multiple continuum and line processes using the Sobolev approximation. 
    The continuum processes include electron scattering and thermal absorption and emission, whilst the line processes 
    include photon scattering and photon destruction by collisional de-excitation.
    Line processes can only occur at specific points in each shell of the stellar wind,
    whereas continuum processes can occur at any point.
    To decide whether a continuum or line process takes place, we apply a similar approach 
    to Mazzali \& Lucy (\cite{mazzali:lucy}), where a random optical depth value is compared to the combined 
    optical depth for line and continuum processes along a photon's path.
    Our code was improved in the following way: after it is decided that the next process will be a continuum process, a 
    second random number is generated to decide which continuum process takes place, an electron scattering or absorption 
    process (as implemented in Vink et al. \cite{vink:dekoter:lamers}; Vink \cite{vink}).
     
    The radiative acceleration of the wind is calculated by following 
    the fate of the large number of photons   
    where the atmosphere is divided into a large number of concentric
    thin shells with radius $r$ and thickness $\ud r$,
    and the loss of photon energy, due to all scatterings that occur within each shell, is determined.
    The total line acceleration per shell summed over all line scatterings in that shell equals 
    (Abbott \& Lucy \cite{abbott:lucy})
     \begin{equation} \label{gline-numerical}
      g_{\rm rad}^{\rm line}\,(r) = - \frac{1}{\dot{M}} \frac{\ud L\,(r)}{\ud r} \, ,
     \end{equation}
    where $-\ud L\,(r)$ is the rate at which the radiation field loses energy 
    by the transfer of momentum of the photons to the ions of the wind per time interval.
    
    The line list that is used for the MC calculations consists of over $10^5$ of the strongest lines of
    the elements from H to Zn from a line list constructed by Kurucz (\cite{kurucz}).
    Lines in the wavelength region between 50 and 10000 \AA\ are included with ionisation stages up to stage VII.
    The number of photon packets distributed over 
    the spectrum in our wind model, followed from the lower boundary of the atmosphere, is $2\times{}10^7$.
    The wind is divided into 90 spherical shells with a large number of narrow shells in the subsonic region 
    and wider shells in the supersonic range.

    \subsection{Computing the mass-loss rate for known stellar and wind parameters}  \label{sect-num-comp-mdot}
   
    By neglecting the pressure term
    and using the expression for the line acceleration per shell (Eq.~\ref{gline-numerical}),
    an integration of the Eq. of motion~(\ref{Eq-of-mot1-nd}), 
    from stellar radius to infinity, yields (cf. Abbott \& Lucy \cite{abbott:lucy})
     \begin{displaymath}
    \frac{1}{2}\, \dot{M}\, \left( v_{\infty}^2 + v_{\rm esc}^2 \right) = \Delta L \, ,
     \end{displaymath}
   or equivalently,
    \begin{equation} \label{Mdot-numerical}
     \dot{M} = \frac{2\, \Delta\, L}{v_{\infty}^2 + v_{\rm esc}^2}\, ,
     \end{equation}
   where $\Delta{}L$ is the total amount of radiative energy extracted per second, summed over all the shells.
   This equation is now fundamental for determining mass-loss rates numerically
   from the total removed radiative luminosity, for the prespecified stellar and wind 
   parameters $v_{\rm esc}$ and $v_{\infty}$, respectively.

   \subsection{The iteration method: Determination of $\dot{M}$ and the wind parameters} \label{sect-it-method}

\begin{figure}
\centerline{\hbox{\hspace{0cm}\epsfxsize=9cm \epsfbox{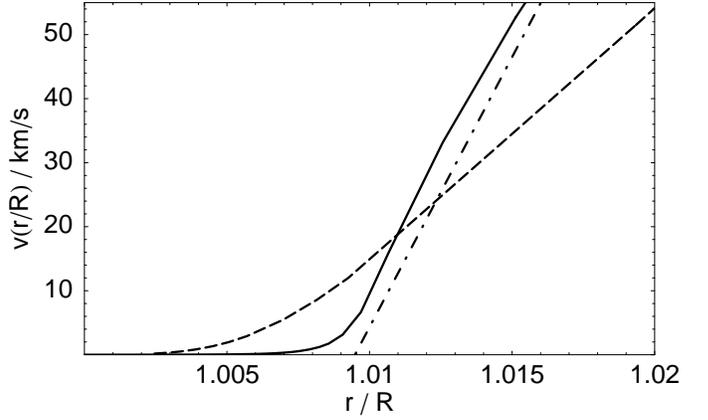}}}
\caption{The numerically computed wind velocity 
         (assumed by {\sc ISA-Wind}, see solid line)
         in comparison to the exact wind solution (dashed line) and the
         $\beta$ wind velocity curve (dotted-dashed line), also used in {\sc ISA-Wind},
         vs. radial distance ${\hat r}$ (in units of $R=11.757\,R_{\sun}$)
         close to the O5--V central star (in Sect.~\ref{sect-applic-ostar}).
          The \emph{exact} solution fulfills the Eq. of motion~(\ref{Eq-of-mot3}) for the converged parameter values
         ${\hat g_{0}}=17661$, $\gamma=0.4758$, $\delta{}=0.6878$ and ${\hat r_{0}}=1.0016$ 
         (shown in the last row of Table~\ref{tab1-itv_16}),
          at the end of the iteration process described in Sect.~\ref{sect-it-method}.
         In this case of convergence, 
         the critical radius ${\hat r}_{\rm c}$ of the Eq. of motion~(\ref{Eq-of-mot3}), determined by Eq.~(\ref{crit-rad-eq}),
         becomes equal to the sonic radius ${\hat r}_{\rm s}=1.011$ determined by {\sc ISA-Wind} (as demanded by our theory
         and iteration method).
         At this radius ${\hat r}=1.011$, both wind velocity curves, i.e. the exact and the computed velocity curve by {\sc ISA-Wind},	  
         intersect each other.}
\label{pic-v-code-ex-beta}
\end{figure}
    In computing the mass-loss rate and the wind model parameters from a given central star 
    with fixed stellar parameters $L$, $T_{\rm eff}$, $R$, $M$, $\Gamma$,
    the following iterative procedure is applied:
    \begin{enumerate}
     \item By keeping the stellar and wind parameters $\dot{M}_{n}$, $v_{{\infty}_{n}}$, $\beta_{n}$ 
           variable throughout our iteration process, 
	   we use arbitrary (but reasonable) starting values $\dot{M}_{-1}$, $v_{{\infty}_{-1}}$, $\beta_{-1}$  
           in iteration step $n=-1$ (cf. Tables~\ref{tab1-itv_16} -- \ref{tab1-itv_16c}).
     \item For these input parameters, a model atmosphere is calculated with  {\sc ISA-Wind}.
           The code yields the thermal structure, the ionisation and excitation structure, and the population of
	   energy levels of all relevant ions.
           Then, the radiative acceleration $g_{\rm rad}^{\rm line}\,(r_{i})$ is calculated
           for discrete radial grid points $r_{i}$
	   with {\sc MC-Wind} and Eq.~(\ref{gline-numerical}). 
	   In addition, an improved estimate for the mass-loss rate $\dot{M}_{n}^{\rm out}$ is obtained by Eq.~(\ref{Mdot-numerical}),
	   which can be used as a new input value for the next iteration step.
	   Moreover, one obtains a new output value for the sonic radius ${\hat r}_{{\rm s}_{n}}$
           (which has to be equal to the critical radius ${\hat r}_{\rm c}$ of our wind theory).
     \item To determine the improved line acceleration parameters $\gamma_{n}$ (or equivalently $\beta_{n}$),
           ${\delta}_{n}$ and ${\hat r_{{0}_{n}}}$,
	   Eq.~(\ref{line-acc-term3a}) or~(\ref{line-acc-term3b}) is used as the fitting formula to apply 
           to the numerical
	   results for $g_{\rm rad}^{\rm line}\,(r_{i})$, cf. Fig.~\ref{pic-gline-fit}.
     \item By applying Eq.~(\ref{v-inf-rc}) and inserting the current values of parameters $\gamma_{n}$, 	   
	    ${\delta}_{n}$ and ${\hat r_{{0}_{n}}}$, as well as the current sonic radius ${\hat r}_{{\rm s}_{n}}$ for ${\hat r}_{\rm c}$,
	    we obtain a new approximation of the terminal velocity $v_{{\infty}_{n}}$, i.e.
	  \begin{equation} \label{v-inf-next}
      \frac{v_{{\infty}_{n}}}{a}\!\! = \!\! \sqrt{\frac{2}{\hat r_{{0}_{n}}}\! \left( 
           \left( \frac{ {\hat r}_{{\rm s}_{n}}^{{\delta}_{n}} }{ {\hat r}_{{\rm s}_{n}}^{{\delta}_{n}} - {\hat r_{{0}_{n}}}}\right)^{\gamma_{n}} 
	   \!\!\!\! \frac{ {\hat r}_{{\rm s}_{n}}^{{\delta_{n}} - 1} }{\delta_{n} \left( 1+\gamma_{n}\right)} \left( {\hat v}_{\rm crit}^{2} 
	           - 2\, {\hat r}_{{\rm s}_{n}}\right) - {\hat v}_{\rm crit}^{2} {\hat r_{{0}_{n}}}^{1 - 1/ {\delta_{n}}} \right)} \, .
          \end{equation}
     \item With these improved estimates of $\dot{M}_{n}$, $v_{{\infty}_{n}}$, $\beta_{n}$ as new input parameters, 
	   the whole iteration step, defined by items 2 -- 4, is repeated until convergence is achieved.	   
    \end{enumerate}

    \subsection{Physical explanation of the iteration method for calculating the unique wind solution} \label{sect-expl-of-it-meth}
    
    In {\sc ISA-Wind} the equation of motion is not solved explicitly for the supersonic wind region
    and, additionally, under the assumption of any input values of the stellar and wind parameters
    $\dot{M}$, $v_{\infty}$ and $\beta$, any arbitrary 'solution' for the wind velocity field is conceivable.
    However, as from the theoretical point of view, it must be ${\hat r_{\rm c}}\equiv{}{\hat r_{\rm s}}$. 
    Hence, to find numerically the unique wind solution by application of {\sc ISA-Wind},
    Eq.~(\ref{v-inf-next}) guarantees that \emph{at the end} of the iteration process
    this condition ${\hat r_{\rm c}}\equiv{}{\hat r_{\rm s}}$ is just fulfilled 
    for the sonic radius ${\hat r_{\rm s}}$ 
    in {\sc ISA-Wind} for the converged set of wind parameters. 

    Eq.~(\ref{v-inf-next}) follows from this condition for the critical radius inserted into Eq.~(\ref{crit-rad-eq}).
    Since the line acceleration parameter ${\hat g_{0}}$ can be expressed by means of the remaining
    parameters ($\gamma$, $\delta$, ${\hat r_{0}}$) via the terminal velocity $v_{\infty}$
    (cf. Sects.~\ref{sect-interpret-eq-of-critrad} and~\ref{sect-reason-for-approxsol}),
    Eq.~(\ref{vinf-law}) has additionally been engaged for its derivation. 

    Thus, Eq.~(\ref{v-inf-next})
    determines the terminal velocity $v_{{\infty}_{n}}$ that our exact wind solution of
    the EOM~(\ref{Eq-of-mot3}) would need to have, in order to pass through the critical point at radius
    ${\hat r_{{\rm c}_{n}}}={\hat r_{{\rm s}_{n}}}$ for the 
    parameters of the fit to the numerical line acceleration curve obtained previously in the same iteration step
    ($\gamma_{n}$, $\delta_{n}$, ${\hat r_{{0}_{n}}}$).
    This improved estimate of $v_{{\infty}_{n}}$ is then used as a new input parameter in {\sc ISA-Wind} for the terminal velocity
    in the next iteration step.

    Naturally, the (current) position of the critical point is also affected by the parameter value of $\dot{M}_{n}$,
    since the value of $\dot{M}$ is simultaneously being determined iteratively and numerically together with the
    other remaining wind parameters by {\sc MC-Wind} and {\sc ISA-Wind} (with its assumed velocity- and density field)
    in our iteration process.
    By the improved estimate of $\dot{M}$ at each iteration step and the use of {\sc ISA-Wind},
    the lower boundary wind velocity $v_{\rm in}$ (at the inner boundary radius)
    is also gradually being adapted to an improved value
    (via the continuity Eq.~(\ref{Mdot}) and the given inner boundary density).

    Finally, at the end of the iteration process in the case of the convergence of all wind parameters,  
    both different wind velocity curves (i.e. the exact solution of EOM~(\ref{Eq-of-mot3}) and the wind curve computed by {\sc ISA-Wind})
    share the same sonic point, i.e. intersect each other at radius ${\hat r_{\rm s}}$ 
    (see Fig.~\ref{pic-v-code-ex-beta}),
    in order to approach the same terminal velocity value $v_{\infty}$ at large distances from the central star
    as ${\hat r}\rightarrow{}\infty$
    (cf. Fig.~\ref{pic-v-approx-beta}). 
   
    Then, our exact wind solution, analytically expressed by Eqs.~(\ref{vr-windsol}) and~(\ref{x-windsol}),
    fulfills the EOM~(\ref{Eq-of-mot3}) for the converged wind parameters 
    ${\hat g_{0}}$ (or $v_{\infty}$), $\gamma$, $\delta$, ${\hat r_{0}}$.
    It now represents the unique wind solution through the critical point of the EOM~(\ref{Eq-of-mot3})
    for the wind from a given star with fixed stellar parameters $L$, $T_{\rm eff}$, $R$, $M$, $\Gamma$.

   \subsection{The model assumptions and velocity field in the atmosphere code ISA-Wind}

    The {\sc ISA-Wind} code has been described in detail by de Koter et al. (\cite{dekoter:schmutz}, \cite{dekoter:lamers}).
    Here, we will only discuss those model assumptions in {\sc ISA-Wind} which affect our wind formalism
    by using the code in our iteration process.
    
    In {\sc ISA-Wind} the wind velocity structure is divided into two regions:
    \begin{enumerate}
    \item The low-velocity subsonic region determined by an EOM, similar to our Eq.~(\ref{Eq-of-mot3}), \emph{however} in contrast to this,
          neglecting the line acceleration term, and, with an additional term $\propto{}\ud{}T/\ud{}r$ on the rhs of the Eq. 
	  (assuming a non-isothermal wind), and
    \item The high-velocity supersonic region where the $\beta$ velocity law in Eq.~(\ref{beta-law}) is assumed to represent
          the wind solution of the EOM (i.e. including then the line radiation forces).  
    \end{enumerate}
    These two regions are connected at a radius ${\hat r}_{\rm con}$, which is determined, together with the parameter 
    ${\hat r}_{0}'$ (in Eq.~\ref{beta-law}), with the condition that the velocity law is continuous in $v$ and $\ud{}v/\ud{}r$
    at this connection radius.
    The value of ${\hat r}_{\rm con}$ is generally located close to and below the sonic point ${\hat r}_{\rm s}$
    (cf., e.g., Fig.~\ref{pic-v-code-ex-beta},
    where ${\hat r}_{\rm con}=1.010$, ${\hat r}_{0}'=1.0095$
    and ${\hat r}_{\rm s}=1.011$). 
    
    The inner boundary condition of the flow velocity, $v_{\rm in}$ at the inner boundary radius $R_{\rm in}$,
    follows from prescribed (i.e. freely selectable) values of
    the density at the inner boundary condition and mass-loss rate (through the equation of mass continuity in form of Eq.~\ref{Mdot}).
    Typically $R_{\rm in}$ is chosen to be situated at a Rosseland optical depth of $\tau_{\rm R}\gtrsim{}20$.
    Therefore, $R_{\rm in}$ does \emph{not} correspond to the stellar (i.e. phostospheric) radius $R$, that depends significantly on frequency.
    In {\sc ISA-Wind} the photospheric radius $R_{\rm phot}$ is defined as the point where the thermal optical depth at 5555\AA{} equals
    $\tau_{\rm e}=1/\sqrt{3}$.
    In practice is $R_{\rm in}<R_{\rm phot}$.
    
    Hence, $R_{\rm phot}$ follows from the model computation and the chosen model parameters describing the density and velocity structure
    in the photosphere: the radius $R_{\rm in}$ and density $\rho_{\rm in}$ at the inner boundary.
    The wind is specified by the input parameters $\dot{M}$, $v_{\infty}$ and $\beta$.
    The parameters specifying the temperature structure are the luminosity $L$, $R_{\rm in}$ and the minimum temperature $T_{\rm min}$
    (the value below which the wind temperature is not allowed to drop).

    \subsection{The adjustment of the wind formalism to ISA-Wind}  
    
    To be able to apply the analytical wind solution of our EOM~(\ref{Eq-of-mot3}) to model a stellar wind
    from a given central star (with fixed stellar parameters)
    by using {\sc ISA-Wind} to find numerically the unique solution,
    we had to adjust our wind formalism, i.e. our more accurate EOM, to the assumed EOM and wind velocity structure in {\sc ISA-Wind}.
    The EOM~(\ref{Eq-of-mot3}) (and therefore our exact analytical wind solution) considers (allows) a line acceleration
    throughout the whole wind regime, starting above the radius ${\hat r}_{0}^{1/\delta}$,
    whereas the different EOM in {\sc ISA-Wind} is only solved in the subsonic region by neglecting the line force.
    However then, {\sc ISA-Wind} 'switches on' the line force somewhere below the connection radius ${\hat r}_{\rm con}$  
    in the subsonic region by assuming a $\beta$ velocity law above ${\hat r}_{\rm con}$ in the supersonic region.
    
    This inconsistency through the use of {\sc ISA-Wind} (compared to our model assumptions and solutions) 
    has been eliminated by introducing the parameter ${\hat r}_{0}$ into our formalism for which the line radiation force is zero
    at radius ${\hat r}_{0}^{1/\delta}$.
    Then, the final value of ${\hat r}_{0}$, together with the other remaining line acceleration or wind parameters,
    can be determined by fitting Eq.~(\ref{line-acc-term3a}) and the
    iteration procedure. 
        
    Further, since {\sc ISA-Wind} begins its computations already \emph{below} (however close to) the stellar (i.e. photospheric) radius,
    all formulae derived in Sect.~\ref{sect-analytical-theory} have thus been applied with reference to the inner boundary (core) 
    radius $R_{\rm in}$ from where the numerical calculations of the wind model start.
    Therefore, the dimensionless variable of distance ${\hat r}$ here refers to $R=R_{\rm in}$
    (see the wind model for a typical O main-sequence star in Sect.~\ref{sect-applic-ostar} and the associated diagrams in
     Sect.~\ref{sect-analytical-theory}).

    \subsection{The chosen boundary values in the wind model} \label{sect-chosen-bound-values}
     
    In our following numerical wind model   
    the inner boundary radius has been chosen (constant throughout the whole iteration process)
    to be $R_{\rm in}=11.757\, R_{\sun}$, situated at a fixed Rosseland optical depth of $\tau_{\rm R}=23$.
    This corresponds then to a photospheric (stellar) radius of $R_{\rm phot}=11.828\, R_{\sun}$ 
    (defined as where the thermal optical depth is $\tau_{\rm e}=1/\sqrt{3}$)
    and an inner boundary density of $\rho_{\rm in}=1.398\times{}10^{-8}$~g/cm$^{3}$ at $R_{\rm in}$.
    
    This particular chosen fixed value of $\tau_{\rm R}$ (or corresponding $\rho_{\rm in}$)
    at each step of the iteration cycle has no effect on the final wind parameters,
    i.e. in particular on the converged values of $\dot{M}$ and $v_{\infty}$,
    as further additional test iterations for a lower boundary value of $\tau_{\rm R}<23$ (at the same $R_{\rm in}$)
    have shown.
    The reason why the exact value of $\rho_{\rm in}$ is irrelevant, is, because we ensure by our iteration scheme
    that the sonic point, determined by the {\sc ISA-Wind} computations, is at each iteration step $n$ always
    located at the right radius, namely at the present critical radius 
    of our EOM~(\ref{Eq-of-mot3}) and exact analytical solution.
    This current critical radius is valid for all the present estimates of updated wind parameter values 
    obtained from the $g_{\rm rad}^{\rm line}$ curve fitting,
    i.e. in particular for $\dot{M}_{n}$ and $v_{{\infty}_{n}}$  
    (see further explanations in Sects.~\ref{sect-interpret-eq-of-critrad} and~\ref{sect-expl-of-it-meth}).

   \subsection{General aspects of the boundary conditions for the uniqueness of the solutions}
    
    To be able to solve the system of equations involving both the equation of continuity~(\ref{eq-continuity2}) 
    and the momentum equation~(\ref{Eq-of-mot1-nd}) uniquely, one boundary condition in each case is required.
    To solve the momentum equation for the velocity $v\,(r)$, the density variable $\rho$ was first eliminated
    by means of the equation of mass conservation, to obtain the density independent equation of motion.
    This equation could then be solved uniquely for (e.g.) the wind velocity,
    because we demand a continuous and smooth trans-sonic flow through the critical point ${\hat r}_{\rm c}$ of
    Eq.~(\ref{Eq-of-mot3}) for a stellar wind as one boundary condition.
    However, to be able to also solve the equation of continuity~(\ref{eq-continuity2}) uniquely for the mass-loss rate $\dot{M}$ by integration,
    as in Eq.~(\ref{Mdot}), knowledge of the density $\rho\,(r)$ for at least one local point $r$ is required
    as another physical constraint. 
    Thus, the density stratification for a given stellar wind (and therefore $\dot{M}$) cannot be determined by merely using
    Eq.~(\ref{Mdot}) with the $v\,(r)$ solution of the equation of motion. 
    
    In more general terms, it can be shown that all solutions ($v_{\infty}$,$\dot{M}$) lie on a curve, which
    fulfills the equation
    \begin{equation} \label{eq-curve-of-solutions}
     \dot{M}\, v_{\infty}^{q} = \mbox{const}\, , \quad \mbox{with}\quad  q\approx{}2\, ,
    \end{equation}
    since $1/2\,\dot{M}\,v_{\infty}^{2}$
    corresponds to the total wind energy (per time interval, i.e. equal to the kinetic wind energy 
    far away from the potential of the central star).
    Hence, by providing an additional boundary condition, as, e.g., a given observed value for $v_{\infty}$,
    the mass-loss rate would then be uniquely determined by a point on this curve 
    (for which $\dot{M}={\rm const}\,v_{\infty}^{-q}$). 
   
    However, since we are interested in self-consistent solutions, i.e. theoretical predictions of $v\,(r)$ {\em and} $\dot{M}$
    for a certain wind model from a given central star, without any prior knowledge of the terminal velocity $v_{\infty}$,
    we follow the additional constraint that $R$ (i.e. the inner boundary radius $R_{\rm in}$)
    is situated at a fixed chosen optical depth $\tau\,(R)$, or corresponding fixed density $\rho\,(R)$
    (cf.~Sect.~\ref{sect-chosen-bound-values}).
    The latter value is given by Eq.~(\ref{rho}) and the values for $\dot{M}$ and $v\,(R)$, i.e. $v_{\rm in}$.
    
    In our iteration process and by the use of {\sc ISA-Wind} this criterion is always fulfilled at each iteration step.
    It is used to update (adapt) the value of $v\,(R)$ at the inner boundary radius of the velocity assumed in {\sc ISA-Wind},
    by means of the new estimate of $\dot{M}$ obtained by {\sc MC-Wind} 
    (and Eq.~\ref{Mdot-numerical} that depends on the value of $v_{\infty}$).
    After all wind parameters have converged,     
    the final position of the sonic radius ${\hat r}_{\rm s}$,
    i.e. the critical radius ${\hat r}_{\rm c}$ belonging to the point through which the unique final wind solution goes,
    is known.  
    
    Hence, to summarise the mathematical point of view,
    the second necessary boundary condition we have used to be able to solve our system of equations 
    (consisting of Eqs.~\ref{eq-continuity2} and~\ref{Eq-of-mot1-nd}),
    or in other words, to find the unique values of $\dot{M}$ and $v_{\infty}$,
    is to prescribe the location of the critical point to be at the \emph{right} position, 
    i.e. at the radius ${\hat r}_{\rm s}$ of the sonic point.
    Whereas the density $\rho\,(R)$ at the inner boundary radius is kept at a fixed value.

    \section{Application: Results for an O--star}  \label{sect-applic-ostar}

%
    \begin{table}
    \begin{center}
    \caption{The different starting values in the iterations A -- C
             for the wind from an O5--V main-sequence star.$^a$} \label{tab-overview-it}
    \begin{tabular}{cll}
    \hline\hline\noalign{\smallskip}
     Iteration &  $v_{\infty}$ [km/s] & $\log\,\dot{M}$ [$M_{\sun}$/ yr]   \\
    \hline\noalign{\smallskip}  
     A & 2020.0 & -5.5 \\
     B & 6000.0 & -7.0 \\
     C & 2020.0 & -7.0 \\
     \noalign{\smallskip}
     \hline 
    \end{tabular}
    \end{center}
    \begin{flushleft}
    $^a$ For the stellar parameters, see Sect.~\ref{sect-applic-ostar}.
    \end{flushleft}
    \end{table}
%
%

%
    \begin{table*}
    \begin{center}
    \caption{\textbf{Iteration A} for the wind from an O5--V main-sequence star. 
     The variable stellar and wind parameters at each iteration step until convergence.$^a$}
    \label{tab1-itv_16}
     \begin{tabular}{rlllllllll}
   \hline\hline\noalign{\smallskip}
   Step &  $v_{\infty}$ & $\log\,\dot{M}$ & $v_{\infty ,\, \rm fit}$  & $\beta$ &  $\gamma_{\rm fit}$ & $\delta_{\rm fit}$ & ${\hat r}_{0, \rm fit}$
   & ${\hat r}_{0}'$ & ${\hat r}_{\rm s}$  \\
   no. & [km/s] &  [$M_{\sun}$/ yr] & [km/s] & & & & & & \\
    \hline\noalign{\smallskip} 
-1 & 2020 &     -5.500 &   --   &     1.0000 &     --     &     --     &     --     &     --     &      --    \\ 
0  & 6403 &     -5.641 &   2365 &     0.8664 &     0.7329 &     0.4917 &     1.0008 &     1.0066 &     1.0175 \\
1  & 4022 &     -6.194 &   3846 &     0.7981 &     0.5963 &     0.7048 &     0.9999 &     1.0087 &     1.0101 \\
2  & 3146 &     -6.278 &   4106 &     0.7598 &     0.5195 &     0.8036 &     1.0020 &     1.0099 &     1.0113 \\
3  & 3003 &     -6.226 &   3909 &     0.7355 &     0.4711 &     0.7747 &     1.0030 &     1.0101 &     1.0115 \\
4  & 2870 &     -6.181 &   3760 &     0.7186 &     0.4372 &     0.7381 &     1.0027 &     1.0100 &     1.0113 \\
5  & 2827 &     -6.141 &   3684 &     0.7047 &     0.4094 &     0.6882 &     1.0024 &     1.0100 &     1.0111 \\
6  & 2603 &     -6.119 &   3495 &     0.7112 &     0.4224 &     0.7203 &     1.0003 &     1.0100 &     1.0110 \\
7  & 2450 &     -6.078 &   3301 &     0.7352 &     0.4704 &     0.7349 &     0.9940 &     1.0099 &     1.0111 \\
8  & 2652 &     -6.038 &   3124 &     0.7538 &     0.5075 &     0.7290 &     0.9939 &     1.0096 &     1.0112 \\
9  & 3211 &     -6.047 &   3232 &     0.7379 &     0.4758 &     0.6878 &     1.0016 &     1.0095 &     1.0112 \\
    \noalign{\smallskip}
     \hline 
    \end{tabular}
    \end{center}
     \begin{flushleft} 
     $^a$ For the fixed stellar parameters $L$, $T_{\rm eff}$, $R$, $M$, and $\Gamma$, see Sect.~\ref{sect-applic-ostar}.
     The line acceleration parameters $\gamma_{\rm fit}$ (or equivalently $\beta$),
     $\delta_{\rm fit}$ and ${\hat r_{0, \rm fit}}$, and the terminal velocity $v_{\infty ,\,\rm fit}$,
     were determined by the fitting formula Eq.~(\ref{line-acc-term3a}) (or Eq.~\ref{line-acc-term3b}),
     applied to the results from a numerical calculation of the line acceleration ${\hat g_{\rm rad}^{\rm line}}\,({\hat r}_{i})$.
     The parameter ${\hat r}_{0}'$ (in the $\beta$ velocity law, Eq.~\ref{beta-law}) and the sonic radius ${\hat r}_{\rm s}$
     are output values from {\sc ISA-Wind}, whereas $\log\,\dot{M}$ is the improved estimated mass-loss rate numerically obtained
     (by {\sc MC-Wind} and Eq.~\ref{Mdot-numerical}).
     At each iteration step, the value of $v_{\infty}$ was calculated by Eq.~(\ref{v-inf-next})
     and used as the new input value for the terminal velocity in the next iteration step,
     together with the new estimates of $\dot{M}$ and $\beta$ (cf. description of iteration process in Sect.~\ref{sect-it-method}).
     Convergence is achieved when 
     the values of $v_{\infty ,\,\rm fit}$ and $v_{\infty}$ have become equal (i.e. here, close) to each other.
     In this case, the condition that the critical radius ${\hat r}_{\rm c}$ (determined by Eq.~\ref{crit-rad-eq})
     has to be equal to the sonic radius ${\hat r}_{\rm s}=1.011$ (determined by {\sc ISA-Wind})
     is fulfilled.
    \end{flushleft}
    \end{table*}
%
%

%
    \begin{table*}
    \begin{center}
    \caption{\textbf{Iteration B} for the wind from an O5--V main-sequence star.
   The variable stellar and wind parameters at each iteration step until convergence.$^a$} \label{tab1-itv_16b}
    \begin{tabular}{rlllllllll}
   \hline\hline\noalign{\smallskip}
   Step &  $v_{\infty}$ & $\log\,\dot{M}$ & $v_{\infty ,\, \rm fit}$  & $\beta$ &  $\gamma_{\rm fit}$ & $\delta_{\rm fit}$ & ${\hat r}_{0, \rm fit}$
   & ${\hat r}_{0}'$ & ${\hat r}_{\rm s}$  \\
   no. & [km/s] &  [$M_{\sun}$/ yr] & [km/s] & & & & & & \\
    \hline\noalign{\smallskip}
-1 & 6000  &     -7.000 &   --   &     1.0000 &     --     &     --     &     --     &     --     &      --    \\  
0  & 4208  &     -6.863 &   7011 &     1.0058 &     1.0116 &     1.3468 &     0.9976 &     1.0107 &     1.0142 \\
1  & 4396  &     -6.655 &   5511 &     0.9347 &     0.8695 &     1.1506 &     1.0042 &     1.0100 &     1.0153 \\
2  & 3683  &     -6.561 &   4968 &     0.9020 &     0.8040 &     1.1385 &     1.0002 &     1.0101 &     1.0135 \\
3  & 3861  &     -6.436 &   4588 &     0.8275 &     0.6550 &     0.9511 &     1.0051 &     1.0099 &     1.0133 \\
4  & 3291  &     -6.385 &   4432 &     0.7843 &     0.5687 &     0.9136 &     1.0039 &     1.0102 &     1.0121 \\
5  & 3116  &     -6.297 &   4152 &     0.7535 &     0.5070 &     0.8314 &     1.0036 &     1.0102 &     1.0118 \\
6  & 2849  &     -6.230 &   3922 &     0.7416 &     0.4833 &     0.7919 &     1.0014 &     1.0101 &     1.0115 \\
7  & 2716  &     -6.165 &   3596 &     0.7422 &     0.4844 &     0.7875 &     0.9996 &     1.0100 &     1.0114 \\
8  & 3311  &     -6.117 &   3544 &     0.7118 &     0.4237 &     0.6827 &     1.0046 &     1.0098 &     1.0113 \\
9  & 2768  &     -6.163 &   3765 &     0.6948 &     0.3897 &     0.6935 &     1.0031 &     1.0100 &     1.0109 \\
10 & 2321  &     -6.120 &   3460 &     0.7213 &     0.4426 &     0.7561 &     0.9953 &     1.0101 &     1.0110 \\
11 & 2922  &     -6.041 &   3228 &     0.7198 &     0.4396 &     0.6698 &     1.0008 &     1.0097 &     1.0112 \\
12 & 2629  &     -6.080 &   3385 &     0.7286 &     0.4572 &     0.7009 &     0.9968 &     1.0097 &     1.0108 \\
13 & 3022  &     -6.063 &   3358 &     0.7168 &     0.4336 &     0.6633 &     1.0018 &     1.0097 &     1.0111 \\
14 & 2493  &     -6.102 &   3397 &     0.7334 &     0.4669 &     0.7447 &     0.9955 &     1.0098 &     1.0108 \\
15 & 3008  &     -6.056 &   3307 &     0.7225 &     0.4449 &     0.6696 &     1.0013 &     1.0097 &     1.0112 \\
    \noalign{\smallskip}
     \hline 
    \end{tabular}
    \end{center}
    \begin{flushleft} 
     $^a$ See description of parameters, analogous to iteration A, in Table~\ref{tab1-itv_16} above.      
    \end{flushleft}
    \end{table*}
%
%

%
    \begin{table*}
    \begin{center}
    \caption{\textbf{Iteration C} for the wind from an O5--V main-sequence star:
    The variable stellar and wind parameters at each iteration step until convergence.$^a$} \label{tab1-itv_16c}
    \begin{tabular}{rlllllllll}
   \hline\hline\noalign{\smallskip}
   Step &  $v_{\infty}$ & $\log\,\dot{M}$ & $v_{\infty ,\, \rm fit}$  & $\beta$ &  $\gamma_{\rm fit}$ & $\delta_{\rm fit}$ & ${\hat r}_{0, \rm fit}$
   & ${\hat r}_{0}'$ & ${\hat r}_{\rm s}$  \\
   no. & [km/s] &  [$M_{\sun}$/ yr] & [km/s] & & & & & & \\
    \hline\noalign{\smallskip}
-1 & 2020  &    -7.000 &   --   &     1.0000 &     --     &     --     &     --     &     --     &      --    \\  
0  & 4161 &     -6.448 &   4467 &     0.9230 &     0.8459 &     1.0278 &     1.0052 &     1.0097 &     1.0203 \\
1  & 4080 &     -6.430 &   4506 &     0.8463 &     0.6927 &     0.9593 &     1.0043 &     1.0097 &     1.0130 \\
2  & 3431 &     -6.407 &   4515 &     0.7937 &     0.5875 &     0.9191 &     1.0040 &     1.0101 &     1.0121 \\
3  & 3340 &     -6.326 &   4260 &     0.7553 &     0.5107 &     0.8416 &     1.0052 &     1.0102 &     1.0119 \\
4  & 2791 &     -6.272 &   4087 &     0.7300 &     0.4600 &     0.7965 &     1.0025 &     1.0102 &     1.0115 \\
5  & 2619 &     -6.179 &   3683 &     0.7321 &     0.4642 &     0.7760 &     0.9995 &     1.0101 &     1.0114 \\
6  & 2924 &     -6.111 &   3484 &     0.7251 &     0.4501 &     0.7078 &     1.0015 &     1.0099 &     1.0113 \\
7  & 2917 &     -6.116 &   3509 &     0.7098 &     0.4196 &     0.6994 &     1.0028 &     1.0098 &     1.0110 \\
8  & 2752 &     -6.116 &   3560 &     0.7073 &     0.4146 &     0.6899 &     1.0014 &     1.0099 &     1.0110 \\
9  & 2688 &     -6.094 &   3450 &     0.7084 &     0.4169 &     0.6930 &     1.0007 &     1.0099 &     1.0110 \\
10 & 2800 &     -6.076 &   3375 &     0.7052 &     0.4105 &     0.6833 &     1.0019 &     1.0098 &     1.0110 \\
11 & 2828 &     -6.082 &   3438 &     0.6978 &     0.3956 &     0.6657 &     1.0024 &     1.0098 &     1.0109 \\
12 & 2416 &     -6.090 &   3369 &     0.7119 &     0.4238 &     0.7133 &     0.9969 &     1.0099 &     1.0108 \\
13 & 2572 &     -6.039 &   3142 &     0.7394 &     0.4788 &     0.7235 &     0.9950 &     1.0097 &     1.0111 \\
14 & 3087 &     -6.035 &   3181 &     0.7320 &     0.4639 &     0.6859 &     1.0012 &     1.0095 &     1.0111 \\
    \noalign{\smallskip}
     \hline 
    \end{tabular}
    \end{center}
    \begin{flushleft} 
     $^a$ See description of parameters, analogous to iteration A, in Table~\ref{tab1-itv_16} above.      
    \end{flushleft}
    \end{table*}
%
%

%
    \begin{table*}
    \begin{center}
    \caption{The results of the different iterations A--C for the wind from an O5--V main-sequence star.
    The converged values of the wind parameters $v_{\infty}$, $\beta$, $\gamma$, $\delta$ ,${\hat r}_{0}$ in the radiative line acceleration
    Eq.~(\ref{line-acc-term3a}) (or Eq.~\ref{line-acc-term3b}, respectively), and mass-loss rate $\log\,\dot{M}$
    of each iteration cycle.$^a$} \label{tab-results}
    \begin{tabular}{llllllll}
   \hline\hline\noalign{\smallskip}
   Iteration & $\log\,\dot{M}$ & $v_{\infty}$  & $\beta$ &  $\gamma$ & $\delta$ & ${\hat r}_{0}$ & ${\hat r}_{\rm c}$\\
             & [$M_{\sun}$/ yr] & [km/s] & & & & \\
    \hline\noalign{\smallskip}
A & -6.047 & 3232 & 0.7379 & 0.4758 & 0.6878 & 1.0016 & 1.0110 \\
B & -6.056 & 3307 & 0.7225 & 0.4449 & 0.6696 & 1.0013 & 1.0083 \\
C & -6.035 & 3181 & 0.7320 & 0.4639 & 0.6859 & 1.0012 & 1.0101 \\
\hline
\noalign{\smallskip}
Mean values & -6.046 $\pm$ 0.006 & 3240 $\pm$ 37 & 0.731 $\pm$ 0.005 & 0.462 $\pm$ 0.009 & 0.6811 $\pm$ 0.0001 & 
               1.0014 $\pm$ 0.0001 & 1.0098 \\ 
    \noalign{\smallskip}
     \hline 
    \end{tabular}
    \end{center}
     \begin{flushleft} 
     $^a$ In addition, the critical radius ${\hat r}_{\rm c}$ (equal to the sonic radius ${\hat r}_{\rm s}$), 
          determined by Eq.~(\ref{crit-rad-eq}), is displayed in the last column, 
          as it is needed as a further parameter to evaluate the wind solution by Eqs.~(\ref{vr-windsol}) and~(\ref{x-windsol}). 
          The mean parameter values (with error estimations) have been calculated from the results of all iterations A--C.    
    \end{flushleft}
    \end{table*}
\begin{figure*}
\centerline{\hbox{\hspace{0cm}\epsfxsize=9cm \epsfbox{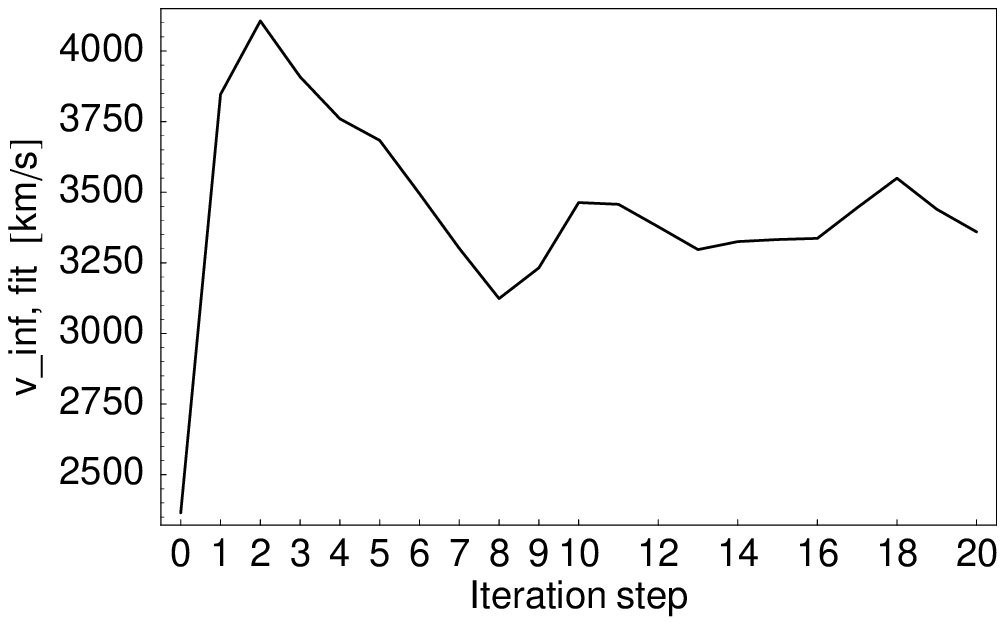}
\epsfxsize=9cm \epsfbox{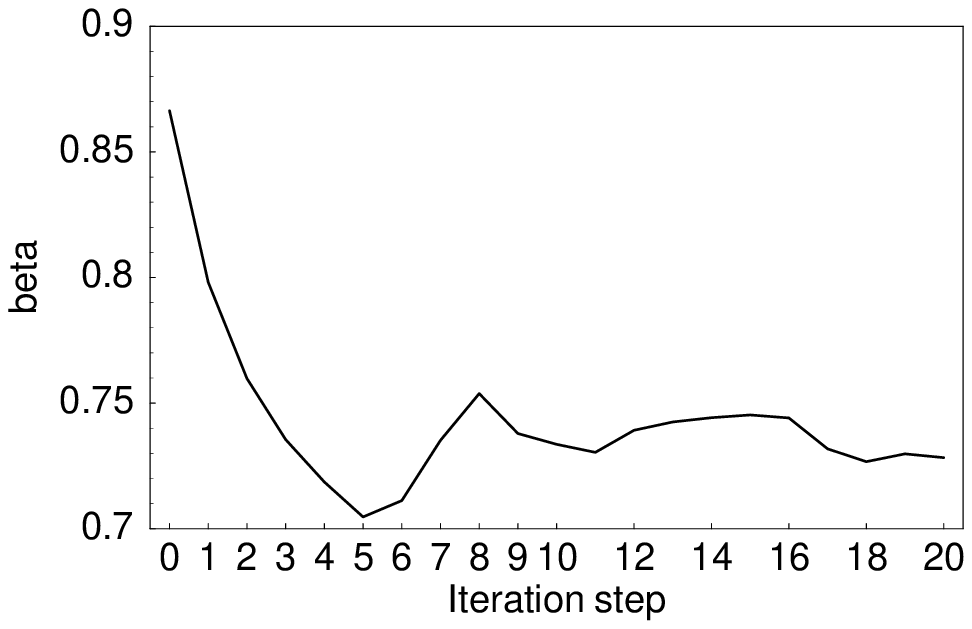}}}
\centerline{\hbox{\hspace{0cm}\epsfxsize=9cm \epsfbox{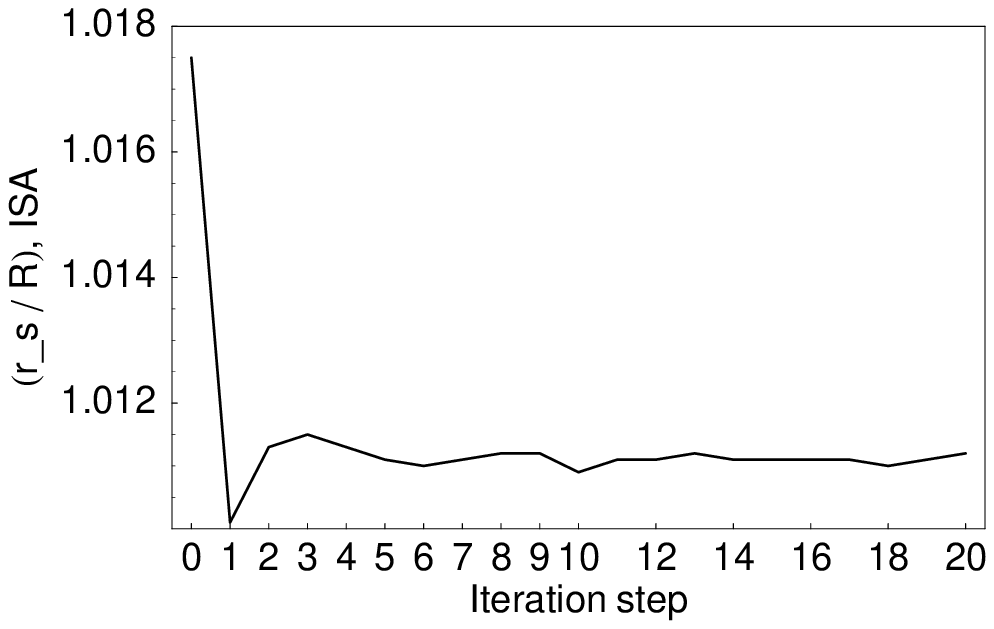}
\epsfxsize=9cm \epsfbox{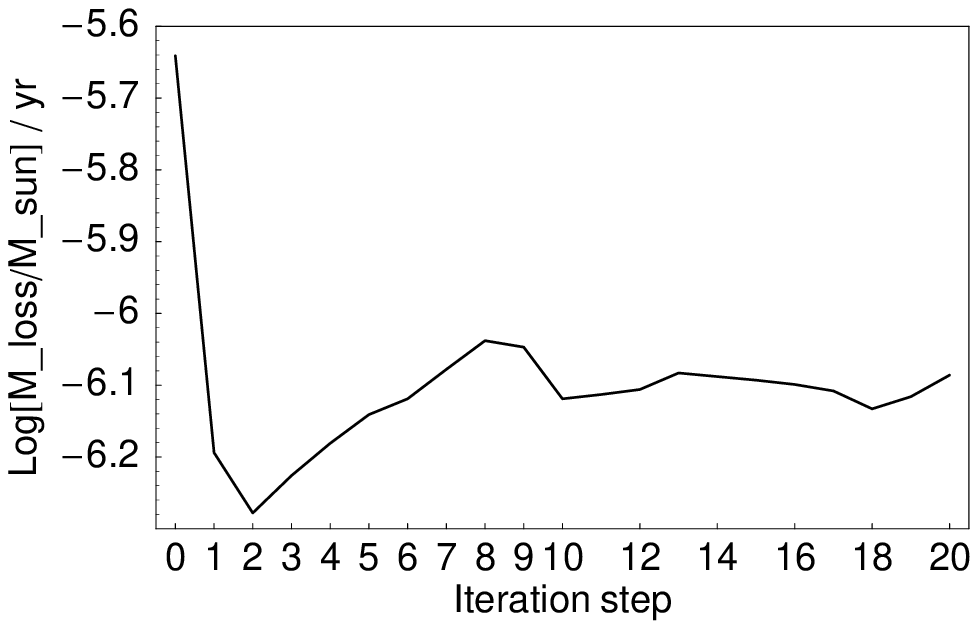}}}
\vfill\parbox[b]{18cm}{\caption[]{\textbf{Iteration A} for the wind from an O5--V main-sequence star: 
      The varying wind parameters $v_{\infty ,\,\rm fit}$ (see upper left diagram) and
      $\beta$ (upper right),
      both obtained by the fitting to numerical results for the line acceleration,
      the numerically determined sonic radius ${\hat r}_{\rm s}$ (by {\sc ISA-Wind}, see lower left),
      and the stellar mass-loss rate $\log{}\dot{M}$ (calculated by MC-Wind, see lower right), 
      plotted versus the step numbers of iteration A until convergence. 
      For the fixed stellar parameters $L$, $T_{\rm eff}$, $R$, $M$ and $\Gamma$, see Sect.~\ref{sect-applic-ostar}.
      See also the description of parameters in Table~\ref{tab1-itv_16}.
      } \label{pic-itv_16} }
\end{figure*}
\begin{figure*}
\centerline{\hbox{\hspace{0cm}\epsfxsize=9cm \epsfbox{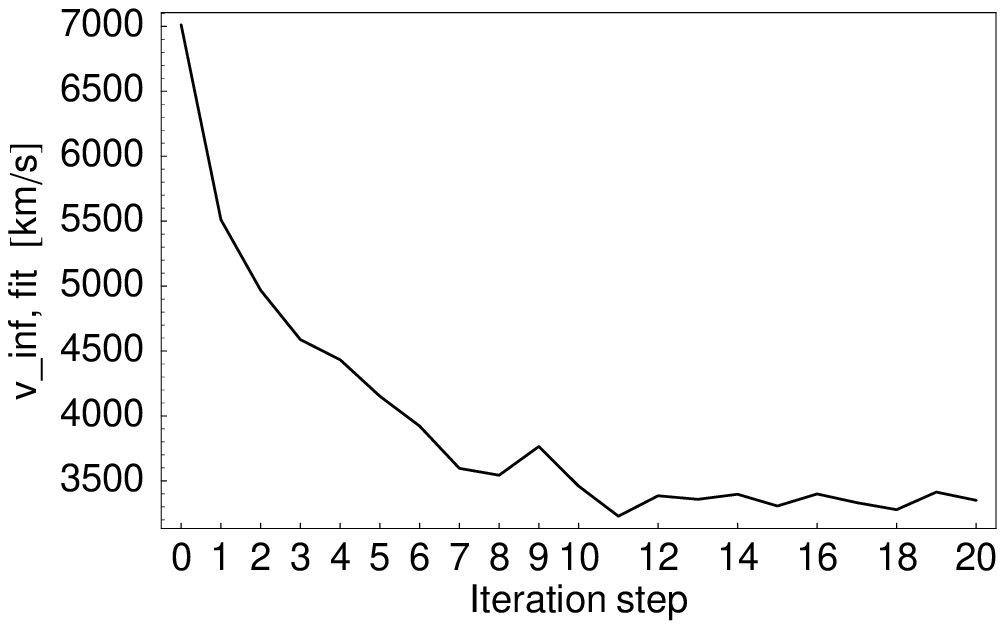}
\epsfxsize=9cm \epsfbox{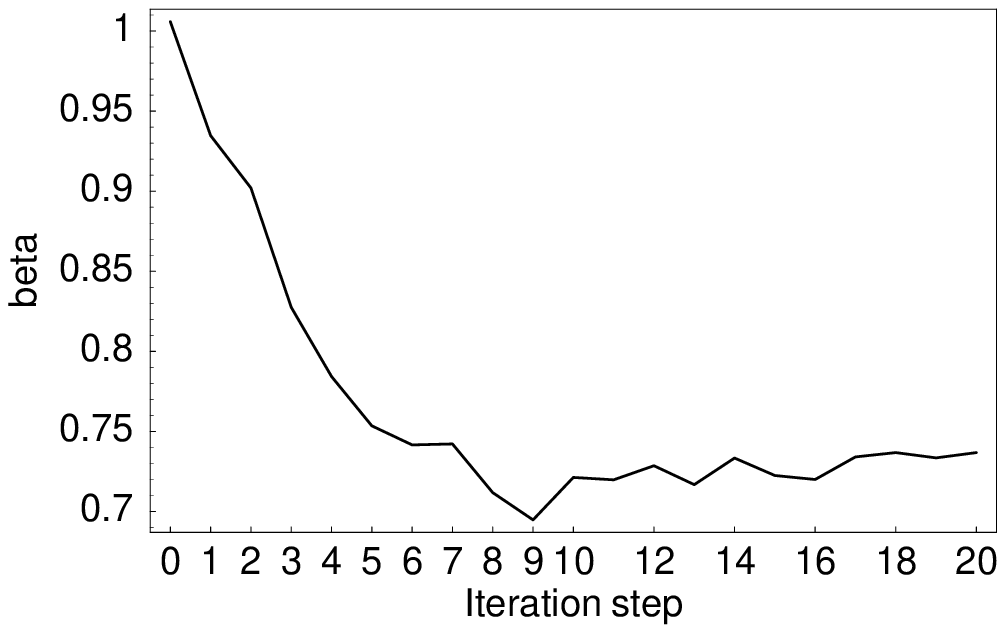}}}
\centerline{\hbox{\hspace{0cm}\epsfxsize=9cm \epsfbox{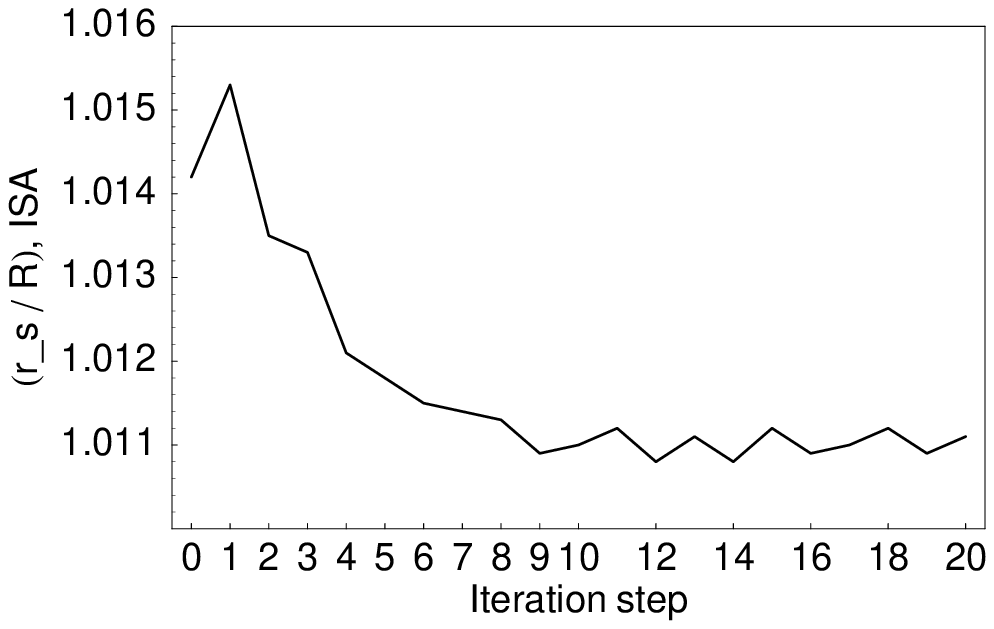}
\epsfxsize=9cm \epsfbox{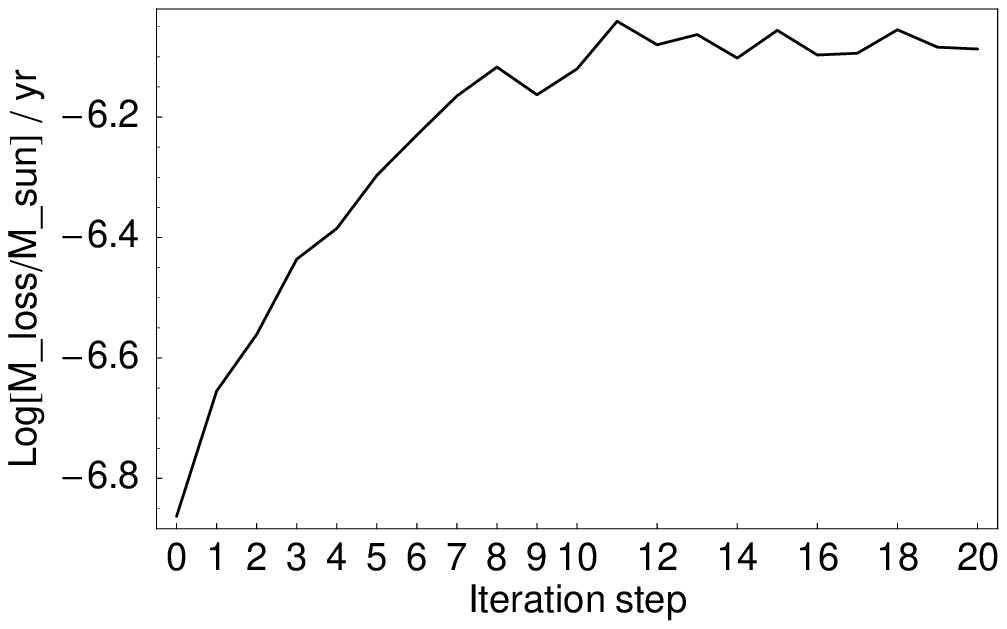}}}
\vfill\parbox[b]{18cm}{\caption[]{\textbf{Iteration B} for the wind from an O5--V main-sequence star:
      The varying stellar and wind parameters  
      $v_{\infty ,\,\rm fit}$ (see upper left diagram),
      $\beta$ (upper right),
      both obtained by the fitting to numerical results for the line acceleration,
      the sonic radius ${\hat r}_{\rm s}$ (determined by {\sc ISA-Wind}, see lower left),
      and the mass-loss rate $\log{}\dot{M}$ (calculated by MC-Wind, see lower right), 
      plotted versus the step numbers of iteration B until convergence. 
      } \label{pic-itv_16b} }
\end{figure*}
\begin{figure*}
\centerline{\hbox{\hspace{0cm}\epsfxsize=9cm \epsfbox{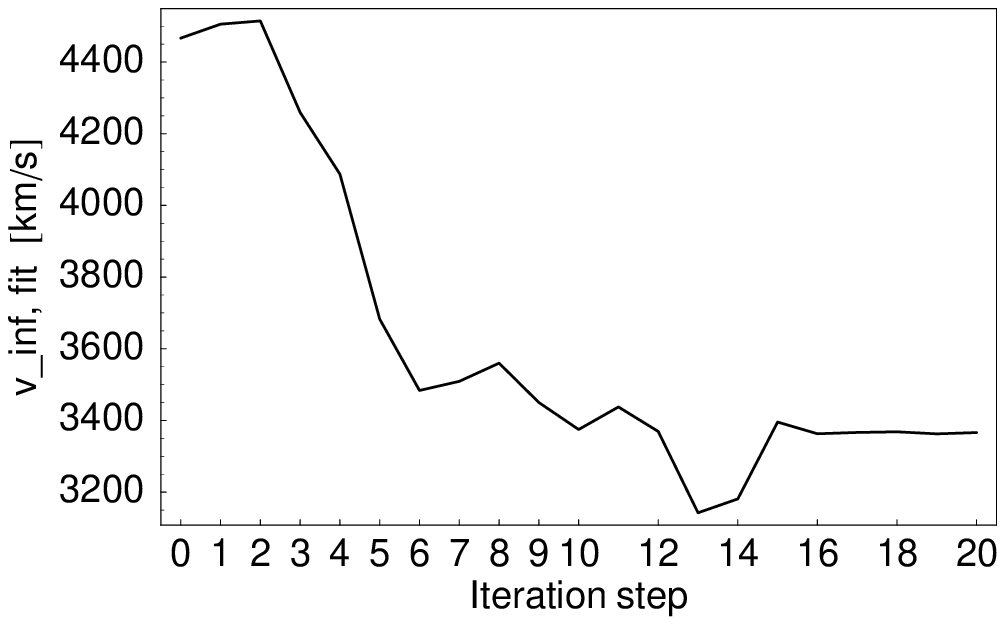}
\epsfxsize=9cm \epsfbox{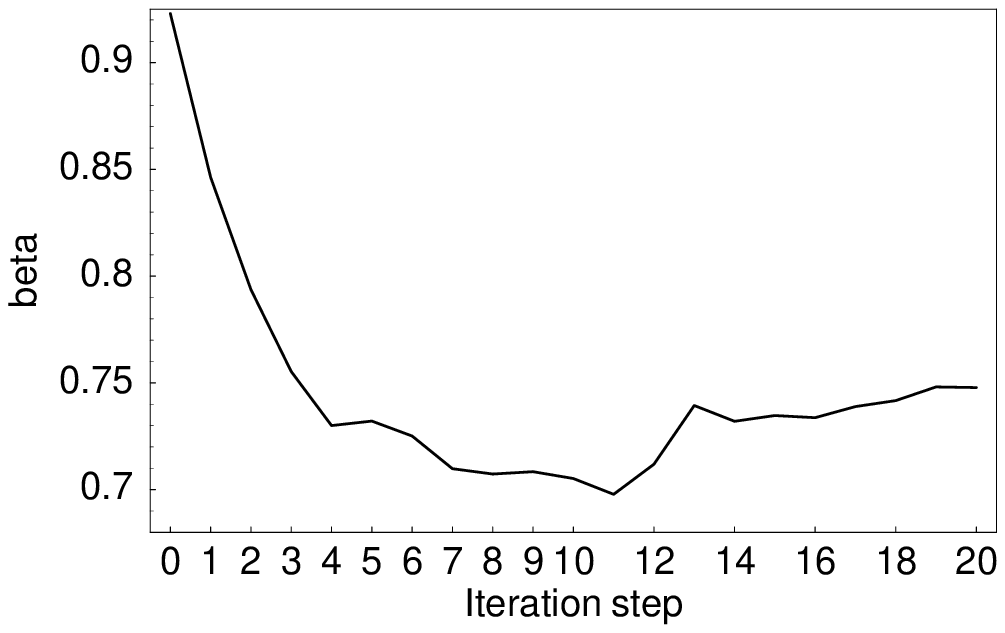}}}
\centerline{\hbox{\hspace{0cm}\epsfxsize=9cm \epsfbox{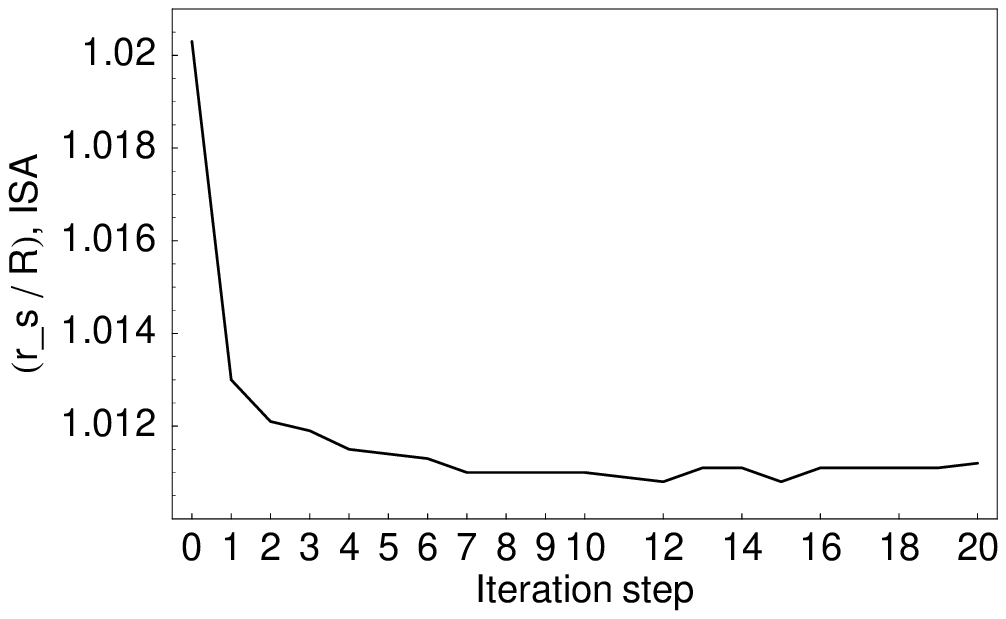}
\epsfxsize=9cm \epsfbox{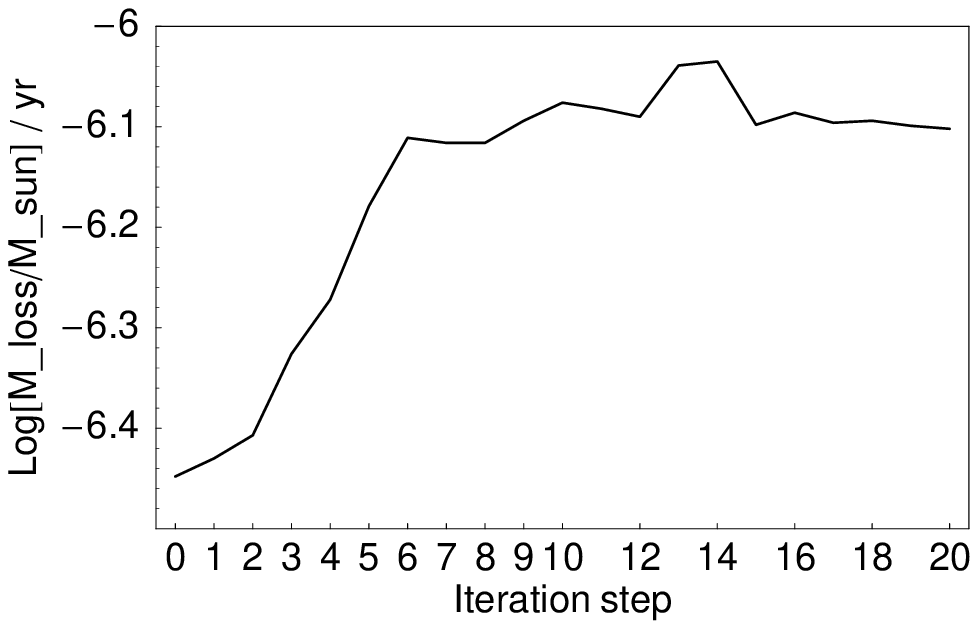}}}
\vfill\parbox[b]{18cm}{\caption[]{\textbf{Iteration C} for the wind from an O5--V main-sequence star:
      The varying stellar and wind parameters  
      $v_{\infty ,\,\rm fit}$ (see upper left diagram),
      $\beta$ (upper right),
      both obtained by the fitting to numerical results for the line acceleration,
      the sonic radius ${\hat r}_{\rm s}$ (determined by {\sc ISA-Wind}, see lower left),
      and the mass-loss rate $\log{}\dot{M}$ (calculated by MC-Wind, see lower right), 
      plotted versus the step numbers of iteration C until convergence.       
       } \label{pic-itv_16c} }
\end{figure*}
      In this section, we apply our theoretical results of Sect.~\ref{sect-analytical-theory}
      and the iterative procedure described in Sect.~\ref{sect-it-method}
      to compute the stellar wind parameters for a typical O5--V main-sequence star. 
      The fixed stellar parameters (see e.g. Martins et al. \cite{martins:al} and references therein) are:

      \begin{itemize}
      \item[-] $T_{\rm eff} = 40000$ K
      \item[-] $\Gamma = 0.214$
      \item[-] $R_{\rm core} = 11.757\, R_{\sun}$
      \item[-] $M = 40.0\, M_{\sun}$
      \item[-] $\log\,\left( L / L_{\sun}\right) = 5.5$
      \item[-] Solar metallicity 
      \end{itemize}    
      From which 
      an isothermal sound speed of 18.16 km/s is obtained.
  
      The iteration method is tested for convergence by three different iteration cycles A--C with three different starting values
      for $v_{\infty}$ and $\dot{M}$, see Table~\ref{tab-overview-it}.
      The reason for choosing these particular values is to start the iteration process for 
      $v_{\infty}$ and $\dot{M}$ values that are either significantly higher, or lower, or lower for both,  
      than the expected values.  

      The results of each iteration step for each of the three iteration cycles are listed 
      in Tables~\ref{tab1-itv_16} to \ref{tab1-itv_16c}. 
      In addition, some of these results are graphically displayed in Fig.~\ref{pic-itv_16} to \ref{pic-itv_16c},
      where the values of $v_{\infty}$, $\beta$, ${\hat r}_{\rm s}$, and $\log \dot{M}$, are plotted against the associated 
      iteration step. 
      
      From this, we can state that for each of the three runs
      all stellar and wind parameters 
      converge to the same values (within statistical fluctuations) within approximately 10 iteration steps.
      However, to achieve a higher precision it is partly necessary to perform up to 15 steps in one iteration
      (see Tables~\ref{tab1-itv_16b} and \ref{tab1-itv_16c} of iteration B and C).
      
      At the end of each iteration process A--C, 
      we obtain the same converged radiative line acceleration ${\hat g_{\rm rad}^{\rm line}}\,({\hat r})$ 
      (see def. Eq.~\ref{def-eq-grad-hat}) of our stellar wind model as, e.g., in case of iteration A,
      plotted versus radial distance ${\hat r}$ in Fig.~\ref{pic-gline-fit}.
      The dots represent the results from the numerical calculation of ${\hat g_{\rm rad}^{\rm line}}\,({\hat r}_{i})$
      for discrete radial grid points ${\hat r}_{i}$.
      These numerical values are fitted well to the line acceleration function in Eq.~(\ref{line-acc-term1})
      by the displayed fitting curve (solid line) with the (fitting, i.e. line) parameters
      ${\hat g_{0}}=17661$, $\gamma=0.4758\pm{}0.0002$, $\delta{}=0.6878\pm{}0.0003$ and ${\hat r_{0}}=1.0016$ 
      (according to $v_{\infty}=3232$~km/s) at the end of iteration cycle A.to find the unique numerical solution
      The remaining converged stellar wind parameters (obtained by the same iteration cycle)
      as well as the resulting parameters from both other iterations B and C
      are listed in Table~\ref{tab-results}. 
      
     Convergence is achieved when both values of the terminal velocity $v_{\infty ,\,\rm fit}$ and $v_{\infty}$, one obtained 
     by the fitting formula Eq.~(\ref{line-acc-term3a}) and the other by Eq.~(\ref{v-inf-next}),
     have approached each other arbitrarily close (up to a desired precision).
     This is the case when the critical radius ${\hat r}_{\rm c}$ (determined by Eq.~\ref{crit-rad-eq})
     has become equal to the sonic radius ${\hat r}_{\rm s}$ (determined by {\sc ISA-Wind}).
      
      Finally, we can summarise the numerical results of all iterations A--C
      by the mean values in Table~\ref{tab-results}:
      $\log\dot{M}=-6.046\pm{}0.006~M_{\sun}/$yr, $v_{\infty}=3240\pm{}37$~km/s, $\beta=0.731\pm{}0.005$,
      or $\gamma=0.462\pm{}0.009$, $\delta=0.6811\pm{}0.0001$ and ${\hat r}_{0}=1.0014\pm{}0.0001$
      with an average critical (i.e. sonic) radius of ${\hat r}_{\rm c}=1.0098$. 
             
      The model results for the wind velocity
      ${\hat v}\,({\hat r})$ as a function of radial distance ${\hat r}$ from the typical O5--V-star,
      with the above mentioned converged wind and line acceleration parameters of iteration A, 
      are displayed in Figs.~\ref{pic-v-approx-ex} -- \ref{pic-v-approx-ex-beta} and have been already discussed
      in Sect.~\ref{sect-analyt-sol-Eq-mot}.
      
      Figure~\ref{pic-v-approx-ex} shows the
      comparison of the approximated and the exact wind solution
      with the converged line acceleration parameters ${\hat g}_{0}$, $\gamma$, $\delta$
      and ${\hat r}_{0}$ of iteration A in Eq.~(\ref{line-acc-term1}),
      where the approximated wind curve was obtained using Eq.~(\ref{vwind-sol-approx}),
      whereas the exact wind solution was obtained from Eqs.~(\ref{vr-windsol}),~(\ref{x-windsol}), and~(\ref{k-branch-wind}),
      with a critical radius of ${\hat r}_{\rm c}={\hat r}_{\rm s}=1.0110$.
      A more detailed comparison of the approximated and the exact wind solution is shown  
      in Fig.~\ref{pic-v-approx-ex-beta}, where both curves are displayed for three different radial ranges.
      The only noticeable discrepancy between the wind velocity curves is in the subsonic and sonic region where the approximated 
      wind solution, derived for (and only supposed to be applied in) the supersonic region, is no longer valid.
      Figure~\ref{pic-v-approx-beta} displays the same approximated wind solution 
      together with the $\beta$ wind velocity curve,
      where the latter was obtained from Eq.~(\ref{beta-law})
      with the converged values for the exponent $\beta=0.7379$ and 
      the terminal velocity $v_{\infty}=3232$~km/s of iteration A (with ${\hat r}_{0}'=1.0095$ as used in {\sc ISA-Wind}).
      Both curves approach each other and the same velocity limit $v_{\infty}$ at large radial distances ${\hat r}$.
      
      These comparisons highlight the 
      good agreement between all velocity curves in the supersonic region,
      except for the noticeable difference of the $\beta$- and approximated wind velocity curve at intermediate distances ${\hat r}\gtrsim{}5$
      from the star (see Fig.~\ref{pic-v-approx-beta}).
      In the supersonic region, the $\beta$ velocity curve always lies (e.g. at radial distance of ${\hat r}=10$ by 12\%, 
      and at ${\hat r}=100$ only by 3\%) above the approximated solution.
      Compared to the exact wind solution, this means that the $\beta$-wind law overestimates the 
      real wind velocity over the entire range of radial distances.  
      
      This difference at intermediate distances is mainly due to our approximation (Eq.~\ref{beta-gam-rel}) for the relationship
      between the exponents $\beta$ and $\gamma$ (cf. the comparison of the $\beta$ law with our approximated velocity)
      that is only approximately independent of the exponent $\delta$, if the latter is close to 1.
      Since in our wind model the value of $\delta$ has actually converged to $0.7$, this approximated relation does not provide a better
      value for $\beta$
      to adapt the $\beta$ curve closer to the approximated (and also real) velocity curve in the middle wind regime.
      Nevertheless, we first had to introduce the additional exponent $\delta$ in our line acceleration function, Eq.~(\ref{line-acc-term1}),
      to gain an optimal fitting to the numerical calculation of ${\hat g_{\rm rad}^{\rm line}}\,({\hat r})$
      at those middle radial ranges of the wind.

     \section{Discussion and conclusions}   \label{sect-disc}

In this paper, we have derived analytical solutions for the velocity structure 
for any mass outflow or inflow situation, through use of the Lambert W-function.
For the case of a radiation-driven wind, we described the line acceleration as a function of 
stellar radius $g(r)$, and therefore found, as a mathematical consequence,
the critical point of our equation of motion to be the sonic point
(in contrast to the critical point of the different EOM of CAK). 

The reason why the line acceleration can be expressed as a function of radius only can be justified as follows:
By solving the basic hydrodynamic equations under the assumption of a stationary (i.e. steady)
spherical flow, we assume that the problem is solvable time-independently and 
all physical quantities, in particular the solution functions, and especially the
line acceleration must be expressable as a function of radius only. This occurs
at the latest after the physical process has reached an equilibrium.
The evidence is given by this work, where, at the end of three different iteration cycles,
$g(r)$ converges to the same function of $r$ for the same stellar wind model.

The numerical calculation of $g(r)$ was performed through Monte Carlo simulations 
and the wind parameters were simultaneously solved in an iterative way for a typical O5--V 
star.
Our computations converged to $\log\dot{M}=-6.05~M_{\sun}/$yr, $v_{\infty}=3240$~km/s, 
$\beta=0.73$ and ${\hat r}_{\rm s}=1.011$ (or expressed equivalently to 
${\hat g}_{0}=17392$, $\gamma=0.46$ and $\delta=0.68$).  

The question is how these self-consistently derived wind parameters compare to 
earlier computations and observational constraints. 
First, when we employ the mass-loss recipe of Vink et al. (2000) using the 
stellar parameters of  $T_{\rm eff} = 40000$ K, $M = 40.0\, M_{\sun}$, and $\log\,\left( L / L_{\sun}\right) = 5.5$, 
for an unspecified terminal wind velocity, so that the recipe uses the standard value of  $v_{\infty}/v_{\rm esc}$ of 2.6, the 
recipe yields a value for the mass-loss rate of $\log \dot{M}=-5.9~M_{\sun}/$yr. 
In other words, our self-consistent computation yields a mass-loss rate in agreement with the Vink et al. recipe.

We now discuss the values that we found for the velocity field. First of all there is the shape of the velocity field, 
$\gamma=0.46$ and $\delta=0.68$, or more widely recognised as $\beta=0.73$. This value still agrees with 
earlier modified CAK calculations (e.g. Pauldrach et al. \cite{pauldrach:al}) as well as empirical 
line profile analyses (e.g. Puls et al. \cite{puls:al:96}, Mokiem et al. \cite{mokiem:al:07a}). 

We finally turn our attention to our derived terminal wind velocity of $v_{\infty}=3240$~km/s. 
This value is still in line with observations, where blue edges of P Cygni 
profiles do not extend beyond $\sim$ $3000$~km/s for O star dwarfs.
We assumed the same
solar abundances as in Vink et al. (\cite{vink:al:00}), which are based on those of Allen (\cite{allen}).
However, a recent 3D analysis of the solar atmosphere by Asplund (\cite{asplund}) suggests that the solar carbon, nitrogen, and oxygen 
abundances may be lower than we have assumed. As these intermediate-mass elements 
are important line drivers in the supersonic portion of the stellar wind (Vink et al. \cite{vink:dekoter:lamers}, Puls et al. \cite{puls:al:00}),
this may have implications for the outer wind acceleration, and hence might somewhat reduce the terminal wind velocity we 
obtain.

Our current 
models are non-rotating, while real O stars rotate rather rapidly. Friend \& Abbott (\cite{friend:abbott}) showed that 
the terminal wind velocity drops with equatorial rotation speed according to 
$v_{\infty}/v_{\rm esc}$ $\propto$ $(1 - \frac{v_{\rm rot}}{v_{\rm break-up}})^{0.35}$. 
As O stars rotate at a significant fraction of their break-up speed, 
our non-rotating models provide just the upper limit to the terminal wind velocity 
(and the lower limit to the mass-loss rate). 
Further investigations, taking into account the effects of rapidly rotating O stars,
predict (up to $\sim$10\%) lower terminal velocities.

The above-mentioned issues will form the basis for future 
investigations, for which we have laid the groundwork 
in the present paper.

      \begin{acknowledgements}
      We thank A.~de Koter for the use of his model atmospheres and R.~Wehrse for helpful discussions. 
      We acknowledge our referees, A.~Feldmeier and S.~Owocki, for their constructive criticism.
      This work was supported by PPARC/STFC.
      \end{acknowledgements}

\end{document}